\documentclass[%
amsmath,
amssymb,
twocolumn,
]{revtex4-1}

\usepackage[a4paper]{geometry} 
\usepackage{tabularx}
\usepackage{array}

\usepackage{graphicx}
\usepackage{dcolumn}
\usepackage{bm}
\usepackage[normalem]{ulem} 
\usepackage[utf8]{inputenc}
\usepackage[T1]{fontenc}
\usepackage{mathptmx}
\usepackage{etoolbox}
\usepackage{color}

\usepackage{seqsplit}

\makeatletter
\def\@email#1#2{%
 \endgroup
 \patchcmd{\titleblock@produce}
  {\frontmatter@RRAPformat}
  {\frontmatter@RRAPformat{\produce@RRAP{*#1\href{mailto:#2}{#2}}}\frontmatter@RRAPformat}
  {}{}
}%
\makeatother

\begin{document}

\preprint{AIP/123-QED}

\title{Electrodrying in nanopores: from fundamentals to iontronic and memristive applications}
\author{Giovanni Di Muccio$^{1,2,\dag}$}
\author{Gon\c{c}alo Paulo$^{1,\dag}$}
\author{Lorenzo Iannetti$^{1,\dag}$}
\author{Adina Sauciuc$^{3}$}
\author{Giovanni Maglia$^3$}
\author{Alberto Giacomello$^{1,\ast}$}
\email{alberto.giacomello@uniroma1.it}
\affiliation{$^1$ Dipartimento di Ingegneria Meccanica e Aerospaziale, Sapienza Universit\`a di Roma, Rome, Italy}
\affiliation{$^2$ NY-MaSBiC, Dipartimento di Scienze della Vita e dell'Ambiente, Universit\`a Politecnica delle Marche, Ancona, Italy}
\affiliation{$^3$ Groningen Biomolecular Sciences \& Biotechnology Institute,  University of Groningen, Groningen, The Netherlands}

\date{\today}

\begin{abstract}
Iontronics is a burgeoning paradigm that employs ions in solution as information carriers for sensing and computing, e.g., in neuromorphic devices. The fundamentally different working principle as compared to electronics requires novel approaches and concepts to control the impedance of nanoscale fluidic circuit elements, such as  nanopores. For instance, previous research has focused on voltage-induced pore wetting as a means to trigger conduction in nanopores. The present study explores the opposite counter-intuitive mechanism: using voltage to dry hydrophobic nanopores and, therefore, to turn off conduction. This “electrodrying” concept affords exquisite, bidirectional control over the conductance of nanopores additionally showing hysteresis in the current-voltage curve that is the fingerprint of memristors. Using an analytical model and free-energy molecular dynamics simulations, we explain the physical mechanism underlying electrodrying and provide clear design criteria for solid-state and biological nanopores with bidirectional control over conductance. The electrical behaviour of electrodrying nanopores shows two unique features: i) the hysteresis loop is shifted from the origin, accounting for the fifth, previously unreported memristor type and ii) negative differential resistance is observed over a broad voltage range in which the non-conductive state is favoured by electrodrying. These properties are demonstrated in a short-term memory task and in an iontronic oscillator circuit to showcase their potential in neuromorphic applications and iontronic devices. Finally, we validate our predictions through experiments on engineered dipolar hydrophobic CytK nanopores, whose voltage-dependent conductance substantiates the electrodrying concept. 
\end{abstract}

\maketitle

\section{Introduction}
There is a century-long history of using voltage as a way to change the wetting behaviour of high surface tension liquids \cite{lippmann1875relations}. Lippmann developed a theory to explain the decrease in the observed contact angle of mercury, i.e., electrowetting, by assuming that the solid-liquid surface tension decreases as the square of the applied voltage; since then, several works have expanded on this theory \cite{Mugele2005,jones2005electromechanical} and used it for manipulating droplets \cite{LEE2002259,cooney2006electrowetting,wikramanayake2020ac} and generating electricity \cite{krupenkin2011, boroujeni2020droplet,riaud2021hydrodynamic}.

Dynamically controlling the wettability of surfaces is a desirable feature for several technological applications involving fluids: both electrocapillarity \cite{Grahame1947} and electrowetting have proved successful in manipulating droplets in microfluidic devices \cite{Zeng2004,Teh2008,Nelson2012,Wu2024}. Most of the work related to electrowetting has focused on either hydrophobic surfaces or microscopic channels, but experimental and computational work has also considered the electrowetting of hydrophobic nanopores \cite{Dzubiella2004,Dzubiella2005,Powell2011,Smirnov2011}; extending this sector is the focus of this work.

Hydrophobic nanopores can be --totally or partially-- dry even when immersed in water. The presence of vapour bubbles typically hinders the transport through nanopores. In ion channels, this phenomenon is known as hydrophobic gating and results in the blockage of ionic currents through the pore  \cite{roth2008bubbles,aryal2015,giacomello2020bubble}: if the pore is dry, due to the formation of a vapour bubble, ions cannot permeate with their hydration shell;  in the wet state, instead, the pore is conductive \cite{zhu2012theory}.
Experiments focusing on hydrophobic nanopores with lengths of several $\mu$m showed that it is possible to wet them using voltages under 5V, thus opening new opportunities to control the ionic conductance of nanopores  by electrowetting \cite{Powell2011,Smirnov2011}. There has also been considerable work exploring how  salt concentration, ionic species, and pH of the solution~\cite{Innes2015,Xiao2016,Polster2022,Polster2020} influence the wetting behaviour of nanopores. Simulations demonstrated that the presence of an electric field has a similar effect to that of increasing the pressure of the system, leading to water intrusion into the nanopore~\cite{Xu2011,Gao2022,Vanzo2014}; in addition,
they showed that the electrostatic behaviour and properties of water confined inside a nanopore may significantly depart from that in the bulk, e.g., confinement can change the dipole response of water \cite{Dzubiella2004,Dzubiella2005, rotenberg2023}. 

Recent developments in iontronic and nanofluidic circuit elements
have led to the flourishing of new methods to tune the conductivity of nanofluidic nanopores for computing applications such as  nanofluidic memristors \cite{Najem2018,robin2021modeling,Xiong2023,ramirez2023synaptical}. Electrowetting shows promise in such applications because it allows rapid establishment of conduction upon wetting of nanopores \cite{paulo2023hydrophobically}. However, a strong limitation of this approach is that electrowetting is known to work only in the direction of increasing the wettability of surfaces, i.e., it can be used only for switching on (but not off) conduction in nanopores, thus preventing reversible and controlled operation. In this work, we propose to exploit intrinsic dipoles within nanopores as a means to reversibly and dynamically control, both in the wetting and drying directions, the water occupancy and thus the conductivity of hydrophobic nanopores.
Previous work has occasionally reported that electrowetting could potentially be non-symmetrical with respect to the applied voltage~\cite{paulo2023hydrophobically,klesse2020}, or that perpendicular electric field could disrupt hydrogen bonds in carbon nanotubes and reduce water occupancy~\cite{Kayal2015}, but a coherent theoretical framework and systematic investigations are still missing. 

Building on the framework developed in previous works~\cite{paulo2023hydrophobically,paulo2024voltage}, we explore the changes in the hydration free-energy landscape of dipolar nanopores through theoretical arguments and molecular dynamics simulations as the external transmembrane voltage is varied. A previously unreported phenomenon is found: voltage can be used to hydrophobically gate a nanopore, i.e., making it less conductive, instead of increasing its conductance as in the usual nanopore electrowetting. We designate this phenomenon as \emph{electrodrying}.

We show that the proposed electrodrying mechanism is rooted in the relationship between the stability of the wet state and the polarization of water molecules within the pore. Traditional electrowetting in nanopores operates by increasing the polarization of water molecules under an applied electric field, which typically lowers the free energy of the wet state and promotes water intrusion into the pore~\cite{trick2017}. This happens because, as a first approximation, the alignment of water dipoles with the electric field minimizes the system's electrostatic potential energy. 
Thermodynamically, while this alignment decreases the system's entropy (creating a more ordered state), the significant reduction in internal energy can result in a net decrease in the free energy, thereby stabilizing the wet state (Supplementary Note 1).

\begin{figure*}[ht]
    \centering
    \includegraphics[width=0.9\textwidth]{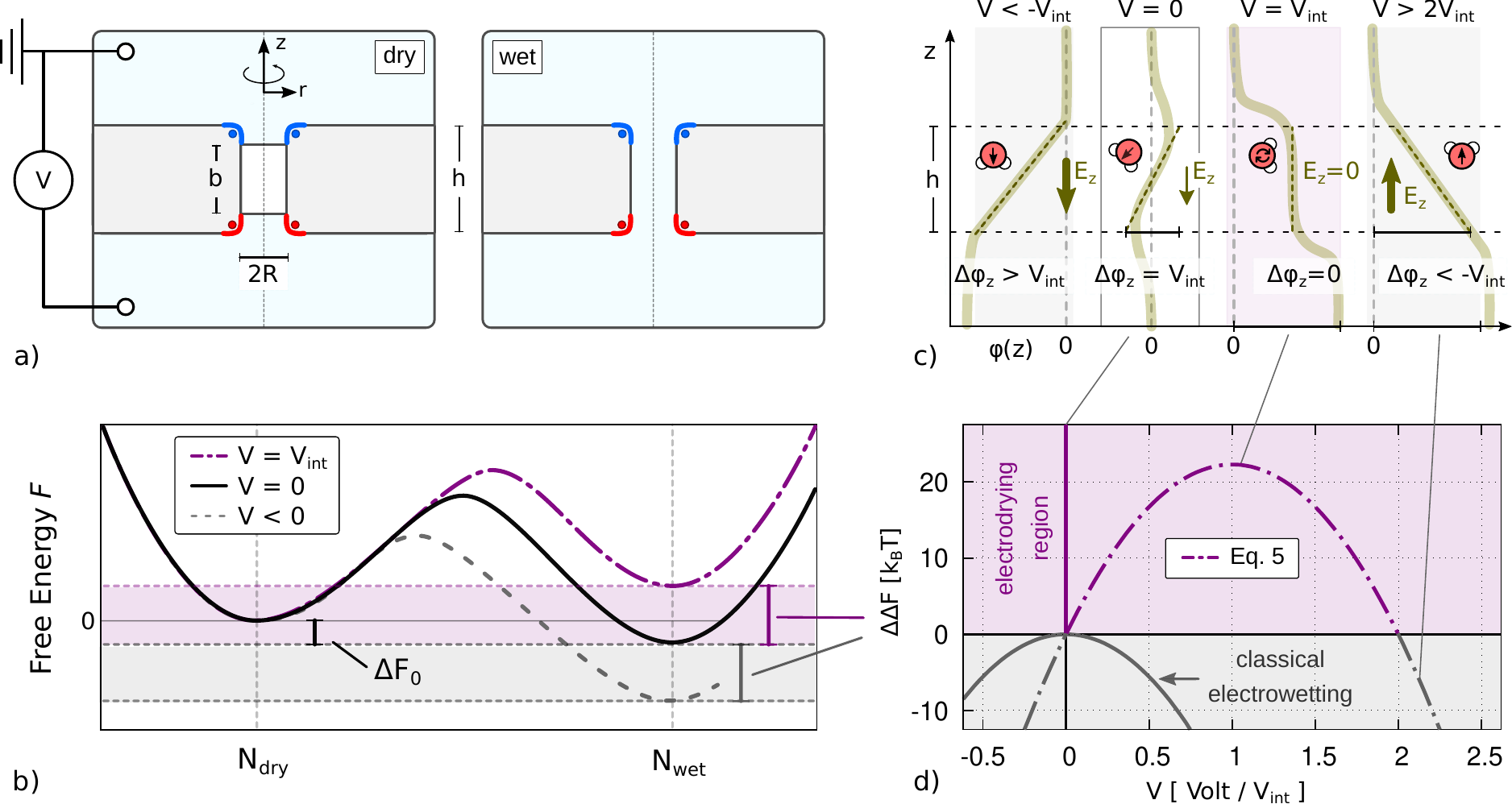}
    \caption{
    \textbf{Electrodrying working principle.}
    \textbf{a)} 
    Schematic of the nanopore system with membrane height $h$, pore radius $R$, and applied potential $V$. Positive and negative charged rings at the pore entrances are indicated in blue and red, respectively. The hydrophobic pore can be either dry (left) or wet (right) due to the presence of a metastable vapour bubble.
    \textbf{b)} 
    Electric potential along the $z$-axis for the wet state. An intrinsic voltage drop 
    $V_{int}$ is present at equilibrium ($V=0$). When an external potential is applied, most of the voltage drop $V$ occurs across the membrane, resulting in an approximated 
    $\Delta \phi_{wet} \approx 
     V_{\text{int}} - V$.
    The electric field $E_z = -\nabla_z \phi(z)$
    varies asymmetrically with $V$. 
    For $V < 0$, $|E_z|$ increases, and for $0< V < 2V_\text{int}$, 
    $|E_z|$ decreases to zero, then increases again.
    \textbf{c)} 
    Hydration free-energy profile as a function of the number of water molecules in the pore. At equilibrium, two local minima correspond to the dry and wet states, with the wet state being more stable ($\Delta F_{wet,0}<0$). For $V < 0$,  the wet state is stabilised, while for $V > 0$, decreased water polarisation leads to a positive variation in $\Delta \Delta F_\mathrm{wet}$, potentially making the pore stably dry.
    \textbf{d)} Hydration free-energy difference between the wet and dry state as a function of $V$, normalised over the intrinsic equilibrium voltage drop $V_\mathrm{int}=1$~Volt, computed via Eqs.~(\ref{eq:dFwet}-\ref{eq:ddF})
    for a pore of length $h=3$~nm and radius $R=0.5$~nm, water relative permittivity $\epsilon_L=80$ and charge of the rings $\pm 7e$.
    }
    \label{fig:system}
    \end{figure*}

If two geometrically identical hydrophobic nanopores are considered, with and without an intrinsic dipole, the one with the intrinsic dipole will favour the wet state.
When an external electric field is applied, depending on its orientation, it may counteract (reinforce) the intrinsic dipole, reducing or even eliminating the polarization of the water molecules within the pore (facilitating electrowetting), as illustrated in Fig.~\ref{fig:system}c.
As polarization is decreased by the external field, the free energy of the wet state increases and destabilises the wet state. Within a voltage range, this competition decreases the stability of the water-filled state, potentially leading to electrodrying of the nanopore. This mechanism can therefore be used to actively control ion conduction in nanopores. 

To test the proposed mechanism, in Sec.~\ref{sec:theory}, we derive an analytical expression to compute the free-energy change between the wet and dry states of hydrophobic nanopores in the presence of an external electric field. In Sec.~\ref{sec:MD}, we present the results of molecular dynamics simulations that validate the theoretical model and generalise it to the case of electrolyte solutions.
Intrinsic dipoles are generated by placing opposite charges at the internal surface of the pore, as often encountered in biological nanopores. 
Results in Sec.~\ref{sec:conductance} show that electrodrying leads to asymmetric, non-linear, and non-monotonic conductivities and to hysteresis in the current-voltage curves, over a range of frequencies. These unique electrical characteristics of electrodrying nanopores correspond to an element with Negative Differential Resistance (NDR)  and to a new memristor type, respectively, that can be leveraged to design bespoke iontronic components. We tested the short-term plasticity of the electrodrying memristor to showcase its use in neuromorphic applications. NDR was harnessed to realize iontronic oscillators.
Finally, in Sec.~\ref{sec:CytK}, the obtained guidelines are leveraged to engineer a biological pore, a CytK mutant, that undergoes electrodrying in the operational range of voltages.

\section{Electrodrying: Working Principle and Theoretical Framework}
\label{sec:theory}

Here, we derive an analytical expression to estimate how the free energy of the wet (conductive) and dry (nonconductive) states of a  hydrophobic dipolar nanopore changes as a function of an external transmembrane voltage. The intrinsic dipole of the nanopore is induced by the presence of oppositely charged rings placed at the pore entrances (Fig.~\ref{fig:system}a). This theoretical expression provides the basis for understanding and engineering electrodrying in nanopores.

The hydration properties of a nanopore subject to an external voltage applied across the membrane can be described by the free energy \( F = F(N,V) \) which is a function of the number of water molecules inside the pore \(N\) and of the external voltage \(V\). For nanoscale hydrophobic nanopores, the free energy may exhibit two minima at different values of $N$, corresponding to the wet and dry states \cite{giacomello2020bubble}; we wish to tune the stability of these states by an external voltage $V$.
For simplicity, we assume that the external voltage results in an electric field parallel to the pore axis,
\[
\mathbf{E} = -\nabla{V} = (0, 0, E_z).
\]
To express the change in free energy  \( \delta F|_{N,V_0} = F(N, V_0+\delta V) - F(N, V_0) \) for a given state $N$ in response to a variation of the external electric potential 
\( \delta V = V - V_0 \), Gubbiotti et al.~\cite{paulo2023hydrophobically,Paulo2023} proposed a second-order expansion of the free energy:

\begin{equation}
    \delta F|_{N,V_0} = P_1(N, V_0) \, \delta V 
    + \frac{1}{2} \, P_2(N, V_0) \, \delta V^2
    \; .
    \label{eq:dF}
\end{equation}
The coefficients \( P_1 \) and \( P_2 \) are related to the total polarization  \( Q=1/L_z\sum_i^N q_i z_i \) along the pore axis \(z\), with \( q_i \) and \( z_i \) the charge and the \(z\)-coordinate of the \(i\)-th particle, respectively:
\[
P_1 = -\langle Q \rangle_{N,V_0} \; , 
\quad P_2 = 
-\beta \left(                 \langle Q^2 \rangle_{N, V_0} - \langle Q \rangle^2_{N, V_0}              \right) 
= \frac{\partial P_1}{\partial V}
\; .
\]

When expansion~\eqref{eq:dF} is performed around \( V_0 = 0 \), systems without a dipole (\( Q = 0 \)) exhibit a symmetric decrease in free energy: both positive and negative $\delta V$ favour the wet state, as in the typical electrowetting scenario~\cite{paulo2024voltage,trick2017,Powell2011}. 
Instead,
the change in free energy due to small voltage variations is asymmetric for systems with a non-zero net dipole moment (\(  Q \neq 0 \)), i.e., it depends on the sign of \( \delta V \).  This asymmetry suggests that the probability of the dry state could increase under the application of a certain external voltage range, i.e. the pore could potentially undergo electrodrying.

To test this insight, we consider a model cylindrical nanopore of radius \( R \) excavated in a hydrophobic solid membrane of thickness \( h \) (Fig.~\ref{fig:system}a). The nanopore has two oppositely charged rings at the two entrances. Apart from the two charged rings, the pore is uncharged. The membrane is immersed in water, without ions or free charge carriers. At equilibrium, the pore is preferentially wet but presents a dry metastable state, with the relative probability of the two states dictated by their free energy difference,
\[
n_{\text{wet}}/n_{\text{dry}} \propto \exp{\left( -\frac{\Delta F_0}{k_BT} \right)},
\]
where 
\( \Delta F_0 = F(N_{\text{wet}}, 0) - F(N_{\text{dry}}, 0) \), \( k_B \) is the Boltzmann constant, and \( T \) the absolute temperature (Fig.~\ref{fig:system}b). 

The presence of oppositely charged rings generates an intrinsic electric field in the system. Focusing on the water particles inside the filled (wet) nanopore, the average polarization in this region can be approximated as~\cite{bonthuis2011dielectric}
\begin{equation}
P_1(N_{\text{wet}}, 0) = - \langle Q \rangle_{\text{wet}} = \alpha_{\text{water}} \, \Delta \phi_z \; \;
                \; ,
    \label{eq:Qw}
\end{equation}
where \( \alpha_{\text{water}} = \epsilon_0 (\epsilon_L -1) \pi R^2/h \), \( \epsilon_0 \) is the vacuum permittivity, \( \epsilon_L \) is the relative permittivity of the liquid, and \( \Delta \phi_z = V_{\text{int}} \) represents the voltage drop due to the intrinsic dipole from the charged rings (Fig. 2b, $V = 0$).
Two simplifying assumptions are used: a) the electric potential is constant across every section of the cylindrical nanopore,
\( \phi(r,z) \approx \phi(z) \), and b) \( \epsilon_L \) is constant. In highly confined environments there can be deviations from such assumptions~\cite{schlaich2016water,becker2024interfacial}; more accurate expressions could be derived or computed, but they are expected to change the reported behaviour only quantitatively. A third assumption is that $\epsilon_L$ does not change with the applied voltage. This approximation is generally valid in water for electric fields up to $\text{1 V/nm}$, for which the orientation energy $p_w E_z \approx k_B T$~\cite{kirby2010micro}, 
with $p_w$ representing the dipole moment of a single water molecule (0.6~{e\AA}).

Since the external voltage drop will  occur primarily across the pore (Fig.~\ref{fig:system}c), the average polarization $P_1(N_{wet}, V)$ at a generic voltage $V$ can be calculated using the superposition principle, focusing on the factor that depends on the external voltage: $\Delta \phi_z \approx V_{\text{int}} - V$. 
Combining this expression with Eq.~\eqref{eq:Qw}, and remembering that the coefficient $P_2$ is the derivative of $P_1(N_{wet}, V)$ with respect to $V$, one obtains
\begin{equation}
    P_2(N_{\text{wet}}, 0) = -\alpha_{\text{water}}
    \; \; .
\end{equation}
Using Eq.~\eqref{eq:dF}, the expected variation of the free energy of the wet state for small applied voltages thus becomes:
\begin{equation}
\delta F_{\text{wet}}(N_{\text{wet}}, V) =
    -\frac{1}{2} 
    \alpha_{\text{water}} 
    V
    \left( 
        V - 2 V_{\text{int}}
    \right) \; .
    \label{eq:dFwet}
\end{equation} 

The free-energy variation \( \delta F_{\text{dry}} \) for the dry state in response to \(\delta V\) can be calculated in the same way; this quantity is needed to calculate the probability ratio of the wet and dry states. 
Consider the system shown in the left panel of Fig.~\ref{fig:system}a, 
with a bubble of length $b$ in the centre of the pore
and water occupying the remaining height $h-b$.
The polarization inside the bubble is trivially zero, since vapour molecules are only observed sporadically in such nanoscale region. 
At the same time, the polarization of the liquid occupying $h-b$ will follow an expression similar to Eq.~\eqref{eq:Qw},
$P_1(N_\mathrm{dry},0) \propto \Delta \phi_z (h-b)/h$.
However, for $b \to h$ this contribution vanishes.
So, we can consider for our system that $P_1(N_\mathrm{dry}, 0) \approx 0$ and, consequently 
\begin{equation*}
    \delta F_{\text{dry}}(N_{\text{dry}}, V) \approx 0 \; .
\end{equation*} 

The free-energy difference between the wet and dry states under an applied electric field can be expressed as \( \Delta F (V) = \Delta F_0 + \Delta \Delta F (V)\) (Fig.~\ref{fig:system}b),
with the relative change driven primarily by changes in the wet state:
\begin{equation*}
    \Delta \Delta F(V) = \delta F_{\text{wet}}(N_{\text{wet}}, V)-\delta F_{\text{dry}}(N_{\text{dry}}, V) \approx \delta F_{\text{wet}}(N_{\text{wet}}, V) =
\end{equation*}
\begin{equation}
   = -\epsilon_0 (\epsilon_L -1) \frac{\pi R^2}{h}
    \frac{V}{2} 
    \left( 
        V - 2 V_{\text{int}}
    \right)
    \; .   \label{eq:ddF} 
\end{equation}
This result, illustrated in Figs.~\ref{fig:system}c-d, reveals an asymmetric response of the system's hydration properties under applied voltage, with the free energy change \( \Delta \Delta F \) being a combination of quadratic and linear functions of \( V \). The quadratic term corresponds to the usual symmetric electrowetting effect, which favours the wet state for both positive and negative \( V \). The linear term, instead, arises from the intrinsic dipole; for  $0<V<2 V_\text{int}$, this contribution is positive. In this voltage range, the probability of the dry state increases (is less stable), reaching a maximum at $V = V_\text{int}$. For larger voltages or negative ones, outside this range, the behaviour aligns with classical electrowetting theory, favouring the wet state.

For short and narrow nanopores with intrinsic dipoles, \( \Delta \Delta F \) can become significant---up to \( 20\, k_B T \) for the nanopore in Fig.~\ref{fig:system} with radius \( R = 0.5 \, \text{nm} \), $V_\mathrm{int}=1$~V, and length \( h = 3 \, \text{nm} \). If the initial free energy difference \( \Delta F_0 \) between the wet and dry states is smaller than this value, the applied voltage can invert the system stability, leading to voltage-induced dewetting (Fig.~\ref{fig:system}b, purple dot-dashed line). This result opens exciting possibilities for the bidirectional control of nanoscale fluid dynamics and conduction through electrostatic modulation, paving the way for innovative applications in nanofluidics, iontronics, and beyond.

\begin{figure*}[ht!]
    \centering
    \includegraphics[width=0.8\textwidth]{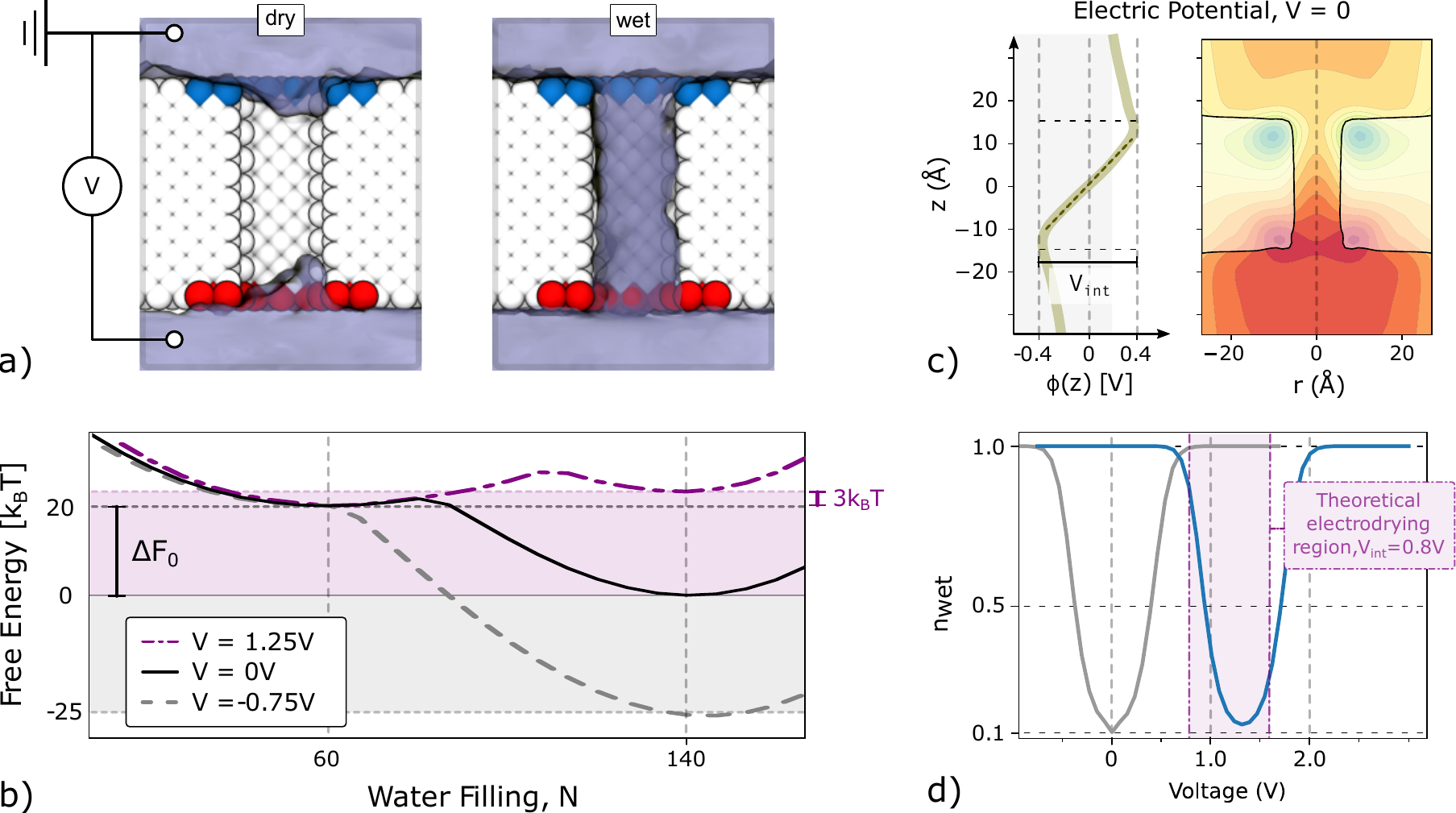}
    \caption{
    \textbf{Simulations of electrodrying in hydrophobic dipolar pores.}
    \textbf{a)} The molecular dynamics system consists of an hydrophobic nanopore with oppositely charged rings at the entrances.
    The top entrance (blue) has a total charge of 
    $+5e$, while the bottom one (red) $-5e$.
    The pore presents a metastable dry state (left) and a stable wet one (right). The pictures are realized using VMD~\cite{vmd}, using the last frame of each state respective RMD simulations; water is represented in transparency.
    A constant and homogeneous electric field 
    $E_z=-V/L_z$ is applied to the system, with $L_z$ the simulation box size in the direction of the pore axis. This setup mimics a potential difference $V$ across the two sides of the membrane.
    \textbf{b)} When no external voltage is applied ($V = 0$), an intrinsic potential difference is present across the pore entrances, $V_\mathrm{int} \approx 0.8\,V$.
    As a positive voltage is applied, this field is cancelled out, eventually drying the pore.
    \textbf{c)} Free energy profiles obtained using RMD. When no voltage is applied the wet state is more favourable. While negative applied voltages increase the stability of the wet state, positive voltages make the dry state more favourable.
    \textbf{d)} Probability to find the system in the wet state for the uncharged hydrophobic nanopore (gray) and for the dipolar pore (blue). The effect of voltage is no longer symmetric around zero, and applying voltage reduces the probability of finding the pore in the wet state to as low as $\sim 10\%$. Increasing the applied voltage wets the pore again. The curve is obtained interpolating the data obtained from eleven free energy profiles computed at different voltages, from $V = -0.75$ to $2.0$~V, using the procedure exposed in~\cite{paulo2024voltage}.
    \label{fig:MD-result}
    }
\end{figure*}

\section{Molecular Dynamics Simulations of Electrodrying}
\label{sec:MD}
\subsection{Dipolar nanopore}
To validate and generalise the theoretical model in Sec.~\ref{sec:theory}, we performed a molecular dynamics (MD) simulation campaign of a model hydrophobic nanopore of length $h=3$~nm and radius $R=0.5$~nm with oppositely charged rings placed at the two entrances, immersed in pure water (Fig.~\ref{fig:MD-result}a).
As a first case, 
we set the total charge of each ring to $\pm 5e$, mimicking a typical charge density present in biological pore (common tetrameric to heptameric assemblies, with charged rings formed by acidic or basic residues, have $\pm 1e$ per residue).
Based on previous calculations on uncharged pores, 
a hydrophobic pore with such dimensions is expected to be dry at ambient pressure conditions ($\Delta F_0 = 15\,\mathrm{k_BT}$) when no voltage is applied across the membrane~\cite{paulo2024voltage,Paulo2023jcp}.
In contrast, the hydration free-energy profile at $V = 0$ of the same pore -- in the presence of the charged dipole -- shows that the stable state is the wet one (Fig.~\ref{fig:MD-result}b, black line), 
with $\Delta F_0 \approx -20\,\mathrm{k_BT}$.
Such significant difference as compared to the completely uncharged case
cannot be explained in terms of a change in hydrophobicity of the system, since the hydrophobic surface surrounding the pore lumen is --in essence-- unchanged.
Rather, due to the charged rings at the pore entrances, an intrinsic electric potential drop ($V_\mathrm{int} \approx 0.8\,V$) is established along the pore axis,
even without any external voltage applied to the system (Fig.~\ref{fig:MD-result}c).
As expected, based on the theoretical framework exposed in Sec.~\ref{sec:theory}, the polarisation of the water molecules induced by the intrinsic dipole stabilises the wet state by more than $20~\mathrm{k_BT}$ as compared to the uncharged case, even when no external voltage is applied~\cite{paulo2023hydrophobically,paulo2024voltage}.

For positive applied voltages $0 < V < 1.25\,V$, 
the free energy of the wet state increases (Fig.~\ref{fig:MD-result}b, purple dot-dashed line). At $V\approx 1.25V$, the free energy of the wet state becomes larger than the free energy of the dry state ($\Delta F \approx 3\mathrm{k_BT}$), thus inverting the wetting behavior of the system (Fig.~\ref{fig:MD-result}d). 
This result is counterintuitive considering the consolidated electrowetting literature, which predicts that voltage always decreases the contact angle of the liquid favoring the wet state. The electrodrying theory in Sec.~\ref{sec:theory} should be invoked, predicting the maximum $\Delta \Delta F$ around $V = V_\mathrm{int} = 0.8$~V.
Further increase of the applied voltage,  
or negative voltages $V < 0$, favour the wetting of the nanopore (Fig.~\ref{fig:MD-result}d), in accordance with theoretical predictions.

With Eq.~\eqref{eq:ddF} in mind, 
and considering that the intrinsic voltage drop 
$V_\mathrm{int}$ depends on the charge difference between the pore mouths, 
we also explore the effect of having less charged pore entrances: $\pm 2.5e$ and $\pm 0.5e$, see Supplementary Fig.~S1. As expected, decreasing the intrinsic dipole across the pore makes the wet state less favourable at $V = 0$, and the electrodrying effect is still present but less conspicuous. In these cases, we still observe a change in the sign of $\Delta \Delta F$ and its asymmetry with respect to $V=0$ but no transition from stable wet to stable dry state is observed, because the pores were already preferably dry at every voltage.

As an additional test to check that the intrinsic dipole  is crucial for observing electrodrying, we simulated a pore in which the atoms at the pore entrances have the same charge (both positive or both negative), i.e., no net intrinsic dipole is present across the pore (Supplementary Fig.~S2). In such cases, the pore is dry when no voltage is applied and, for any $|V| > 0$, the voltage only contributes to (symmetrically) stabilise the wet state of the pore, making the dry state progressively more unfavourable. More analytically, the linear term in Eq.~\eqref{eq:ddF} related to $V_\mathrm{int}$ vanishes, and it is thus not possible to obtain $\Delta \Delta F>0$ (electrodrying).

\subsection{The effect of ions in solution}
The systems considered so far are immersed in pure water, since the analytical expressions derived in Sec.~\ref{sec:theory} are strictly valid only for very diluted systems.
Because we are interested in studying the potential application of electrodrying in iontronics, we check whether electrodrying is possible in the presence of ions.
Theoretically, by adding ions to the solution,
the charged rings can be locally screened at a distance of a couple of Debye lengths $\lambda_D$.
Since $\lambda_D \propto c^{-1/2}$,
as the salt concentration $c$ increases,
the intrinsic voltage drop across the nanopore 
($V_\mathrm{int}$) is expected to decrease, thus reducing water polarization within the nanopore.

Simulations show that, when the nanopore is immersed in a $0.5$~M NaCl solution, the stability of the wet phase is reduced in favour of the dry one at $V = 0$  (Supplementary Fig.~\ref{fig:ions}a). In accordance with electrowetting, the application of an external voltage $V$ increases the stability of the wet state, regardless of the sign, although the effect of opposite voltages $V = \pm 1\,V$ is asymmetric due to the intrinsic dipole (different $\Delta \Delta F$). 
In order to obtain a significant electrodrying effect in the presence of ions, two avenues are possible: i) increasing the intrinsic dipole or ii) lowering the ion concentration. 
Indeed, by decreasing NaCl concentration to 0.05M, the pore is again mainly wet at $V = 0$. Furthermore, the application of a positive $V>0$ favours the dry state (Supplementary Fig.~\ref{fig:ions}b, purple line), leading to a full inversion of stability (electrodrying).
On the other hand, increasing the charge of the two rings to $\pm 10$~e significantly increases the stability of the wet state at $V = 0$.  Applying $V > 0$ increases the stability of the dry state. However, $\Delta \Delta F$ is not large enough to invert the stability of the states, that is, to make the dry state more probable; in this case, electrodrying is obtained only in a weak sense (Supplementary Fig.~\ref{fig:ions}c). 

In summary, there is a spectrum of dipole charges and salt concentrations under which electrodrying can be observed. Such range depends on the characteristics of the pore---size, shape, hydrophobicity (the liquid-solid contact angle or more generally its chemistry)---and on the global charge distribution of the system. In complex scenarios, such as biological confinement, unexpected dipole inversion can occur allowing electrodrying at 1M concentrations, see Sec.~\ref{sec:CytK}.

Electrodrying emerges as a robust phenomenon that can be engineered for a variety of technologically relevant conditions and that should also be relevant for a number of naturally occurring biological pores.

\begin{figure*}[ht]
    \centering
    \includegraphics[width=0.9\textwidth]{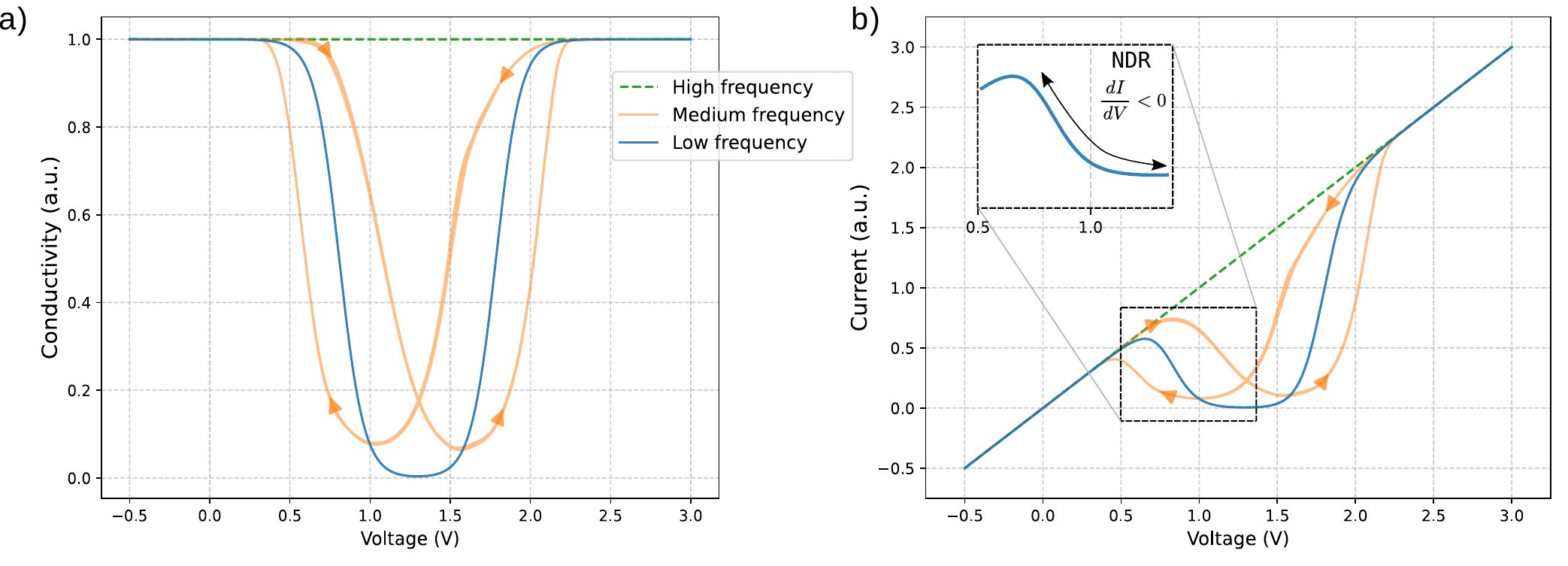}
    \caption{
    \textbf{Atypical electrical characteristics of electrodrying nanopores.} 
    \textbf{a)} Conductance of an array of electrodrying nanopores as a function of voltage. The conductance is computed as $G_o n_\mathrm{wet}(V)$, where $G_o$ is the open-pore conductance and $n_w$ is the probability of the wet (conductive) state. The wetting and drying rates are estimated as the voltage is cycled at different frequencies using the procedure in \cite{paulo2024voltage,paulo2023hydrophobically}. The open-pore conductance is assumed to be constant with voltage and equal to $1$. 
    \textbf{b)} Current as a function of voltage, computed by multiplying the conductance by the voltage.
    Real units for conductance and current can be obtained by multiplying the single pore conductance (usually in the order of 1~nS) at a given voltage for an arbitrary number of pores (tens to hundreds). 
    \label{fig:currents}
    }
\end{figure*}

\section{A New Memristor Type with Negative Differential Resistance}
\label{sec:conductance}

The conductivity of a hydrophobically gated nanopore depends mainly on the probability of it being in a wet state \cite{zhu2012theory,paulo2023hydrophobically}. In the case considered in Fig.~\ref{fig:MD-result}, we can safely assume that the pore is non-conductive when dry and exhibits constant conductivity when wet. This allows us to estimate the conductance of an array of nanopores under voltage cycles with different frequencies. Figures~\ref{fig:currents}a,b show that, at low frequencies, an array of electrodrying nanopores exhibits non-monotonic and non-linear conductance (blue lines). At higher frequencies, the pores exhibit an ohmic behaviour, in which the conductivity becomes independent of the applied voltage (green dashed lines). Interestingly, at intermediate frequencies a peculiar hysteresis loop is observed in the conductivity, with a pinched loop in the current-voltage curve that is the hallmark of memristors~\cite{CHUA2012}. 

The memristive behaviour in Fig.~\ref{fig:currents} is particularly intriguing, constituting a new type of memristor different from the 12 types in the usual classification \cite{sun2019} (Supplementary Fig.~\ref{fig:memristor_types}). Our electrodrying memristor curve crosses the origin of the I-V plane as type 1 (M1-M4) memristors but has a non-zero crossing pinch as type 2 capacitive-coupled memristors at $V \approx 1.25\,V$ and $I\approx 0.25$. This behaviour is different from any reported memristor type~\cite{CHUA2012,sun2019} and nanofluidic implementations thereof~\cite{cervera2006ionic,ramirez2023synaptical,robin2023long,ismail2025programmable}.
We remark that, because the conductance is computed only from the open-pore probability, no capacitive effects are present in Fig.~\ref{fig:currents} thus genuinely accounting for the fifth memristor type.
Additionally, our memristor exhibits prominent negative differential resistance, as discussed in the following.

In traditional unipolar memristors, the pinched hysteresis loop implies a memory effect in which the device conductance depends on the history of applied voltage. The fact that our observed pinch is shifted indicates a deviation from this norm, with memory occurring at $V\neq 0$. Physically, this is due to the presence of an intrinsic dipole that shifts the voltage at which dewetting starts, $V_\mathrm{int}\approx0.8$~V (Fig.~\ref{fig:system}c); the pinch point corresponds to the minimum in the wet probability at $V \approx 1.25\,V$ (Fig.~\ref{fig:MD-result}d). Drawing a parallel with the Hodgkin and Huxley model of the action potential~\cite{hodgkin1952quantitative}, the wet probability, i.e., the probability of having a conductive pore, plays the role of the gating variable that characterise the response of each ion channel. For example, it would be possible to engineer electrodrying memristors with different gating thresholds and dynamics by tuning the pore geometry and the charge at the end of the pores, to mimic the sodium and potassium channels.

The second notable electrical characteristic of our electrodrying nanopore is the emergence of Negative Differential Resistance (NDR) \cite{chua1983negative} both in the static and memristive regimes. NDR is evident in the I-V curves in Fig.~\ref{fig:currents}b, where, between $V_\mathrm{int}$ and the minimum of $n_{wet}$, the current decreases with increasing applied voltage. Our electrodrying nanopore behaves as an N-type NDR element with a single working point at each voltage. More formally, the differential conductance $G_\mathrm{diff} = dI/dV$ becomes negative satisfying: 
\begin{equation}
\label{eq:NDR}
\frac{dI}{dV} =
\frac{dG(V)}{dV} V + G(V) < 0
\implies
-\frac{dG(V)}{dV} > \Big\lvert \frac{G(V)}{V} \Big\lvert \, .
\end{equation}

This peculiar conduction property is a direct consequence of electrodrying that decreases the conductance by opening the hydrophobic gate when the voltage increases ($dG/dV < 0 $). This latter is a necessary condition, but not sufficient, as stated by Eq.~\eqref{eq:NDR}. In our model system, the term $dG/dV$ is larger in modulus than $|G/V|$, satisfying inequality~\eqref{eq:NDR} for NDR. In more physical terms, the current through a conductance is expected to increase by $G\delta V$ in response to a small voltage increment, but this term is outweighed by the drop in current caused by the decrease in conductance due to electrodrying. 
In classical electrowetting, instead, $dG/dV > 0 $ for positive voltages and $dG/dV<0$ for negative ones as can be seen from the grey curve of Fig.\ref{fig:MD-result}d, implying that inequality~\eqref{eq:NDR} can never be satisfied. An electrowetting nanopore thus corresponds to a nonlinear resistor \emph{without} NDR. 

These findings enable the bidirectional control of gating in nanopores by electrodrying. This is achieved by strategically positioning charges in such a way that hydrophobic gating and electrostatic effects are synergistic. Understanding the nuances of electrodrying not only enhances our grasp of the fundamental physics that governs pore conductance in biological nanopores but also opens potential avenues for designing novel devices with tailored memristive and conductive properties as demonstrated in the two selected applications below.


\begin{figure*}[ht]
    \centering
    \includegraphics[width=0.9\textwidth]{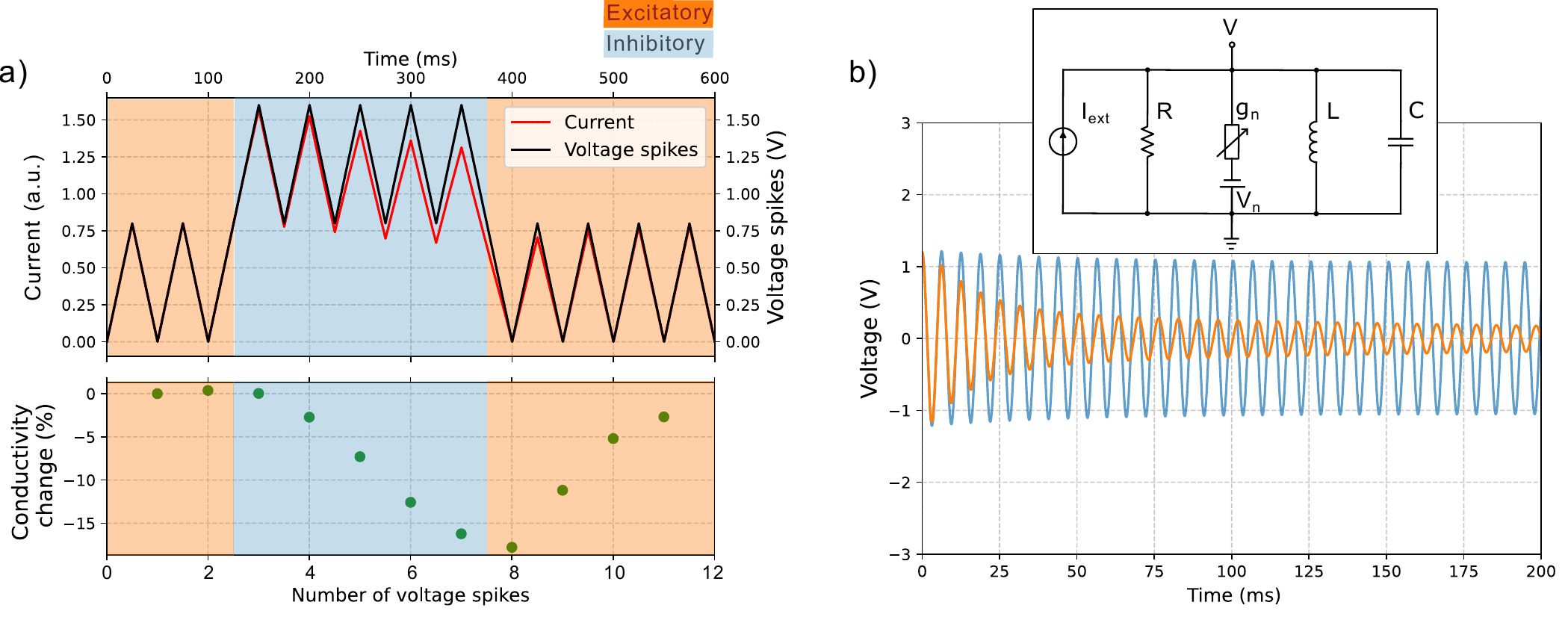}
    \caption{
    \textbf{Memristive and iontronic applications of electrodrying nanopores.} 
    \textbf{a)} Learning-forgetting task on an array of model electrodrying nanopores. A train of voltage spikes composed of two excitatory pulses, five inhibitory pulses, and four excitatory pulses is used to program the conductance of the nanopore array. Current is measured as a function of time (top panel), while memory in the response is estimated as $\Delta G_i(\%) = \left(G_i-G_0\right)/G_0 \cdot 100$ where $G_i$ is the time averaged conductance over a pulse and $G_0$ is the value for the first pulse (bottom panel). Current evolution in time is computed from the wetting and drying rates as explained in \cite{paulo2023hydrophobically}. 
    \textbf{b)} RLC oscillator with electrodrying nanopores. Plots show two different possible oscillating responses of the circuit depending on the model parameters. The blue curve (undamped oscillator) is obtained with an open pore capacitance $g_0 = 100 \ mS/cm^2$ while the inverse of the circuit resistance is $1/R = 1 \ mS/cm^2$. The orange curve (damped oscillator) is produced with $g_0 = 1/R$. The other circuit parameters for the curves are $I_{ext} = 1 \ mA/cm^2$, $L = 1 \ mH/cm^2$, $C = 1 \ mF/cm^2$, $V_n = -1.5 \  V$ 
    \label{fig:applications}
    }
\end{figure*}

The memristive properties of electrodrying nanopores make them an attractive choice for neuromorphic applications. We tested the short-term plasticity of this system by performing a simple learning and forgetting experiment with a procedure similar to Ref.~\cite{paulo2023hydrophobically}. Given the history-dependent response of the system, its conductance can be programmed by sending a train of voltage pulses above and below $V_\mathrm{int}$ that have inhibitory and excitatory effects, respectively (Fig.\ref{fig:applications}a). The nanopore is wet under resting conditions and excitatory pulses do not change its conductance. However, if inhibitory pulses above this threshold are sent, the conductance gradually decreases with each incoming spike, imprinting the memory of the previous pulses in it. These changes in conductance can be successively erased with a sufficient number of pulses below $V_\mathrm{int}$.


Next, we present a classical application of NDR: an undamped RLC oscillator circuit \cite{berger2011chapter5}. 
RLC circuits are electrical systems composed of a resistance R, a capacitance C, and an inductance L connected in series or in parallel, capable of producing oscillations in the output signal due to the resonance between inductance and capacitance. Given a constant external input, the system response will oscillate with a resonance frequency $\omega_0 = 1/\sqrt{LC}$ and will be damped over time due to the nonzero resistance R. When connected in parallel with R, NDR elements introduce positive feedback in the circuit that allows one to cancel out damping.  
We tested our nanopore connecting it in parallel with an RLC oscillator and in series with a voltage generator $V_n$ that shifts the operating point of the electrodrying nanopore close to the NDR range (inset in Fig.~\ref{fig:applications}b). In biological terms, $V_n$ is conceptually similar to the reversal potential of each ion channel of the Hodgkin and Huxley model, which sets the conditions in which a given ion current reverses \cite{hodgkin1952quantitative}.

Figure~\ref{fig:applications}b shows that, by tuning the nanopore conductance, for instance adding more nanopores to the membrane, it is possible to obtain damped or undamped oscillations in response to a DC current. This result demonstrates how the NDR characteristic of electrodrying nanopores can be practically employed.

\section{A bio(techno)logical proof-of-concept}
\label{sec:CytK}
\begin{figure*}[ht!]
    \centering
    \includegraphics[width=1\textwidth]{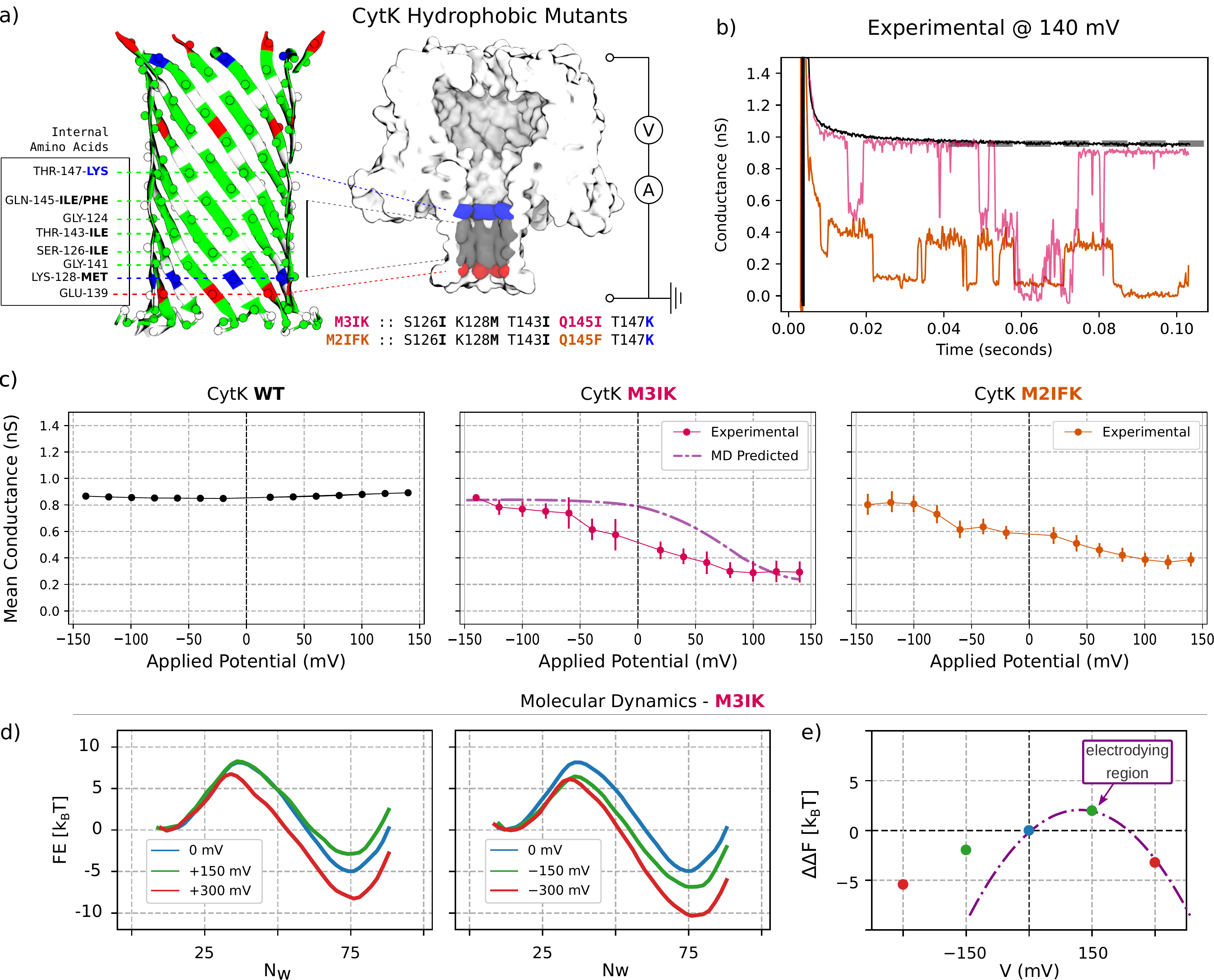}
    \caption{
    \textbf{Engineered dipolar hydrophobic CytK nanopore.} 
    a) Schematic illustration of the transmembrane barrel and 3D renderings of CytK nanopore, showing the spatial arrangement of engineered hydrophobic mutations (gray) alongside charged residues (blue, red) designed to induce electrodrying.
    b) Representative experimental conductance–voltage measurements at 140~mV for the Cytk WT (black), M3IK (pink), and M2IFK (orange) mutants. The presence of pronounced current blockades for the mutants indicate the presence of hydrophobic gating. Dashed lines represents the average conductance between 40 and 100 ms; the voltage is applied at $t=2.5$~ms, see the inital spike.
    c) Mean conductance for different applied voltages, computed by averaging the current traces of >20 voltage sweeps, from two independent experiments per system. Supplementary Fig.~S5 reports additional current traces for other voltages. Overlay of experimental data (symbols) and predictions based on MD results (purple dot-dashed line) for the M3IK mutant confirms that the observed conductance modulation accurately follows the theoretical predictions for hydrophobic electro‑drying.
    d) Free-energy profiles as a function of the water filling $N_w$ of the nanopore, for positive and negative applied voltages, show how the external field alters the barrier and well depths.
    e) Voltage-dependent shifts $\Delta \Delta F$ of the free-energy  difference between minima, using the $V=0$ case as a reference. The parabolic fit (purple dot-dashed line) illustrates the asymmetric electrodrying trend of Eq.~\eqref{eq:ddF} for reference.
    }
    \label{fig:cytk}
\end{figure*}

To showcase electrodrying in a biologically and technologically relevant context, we engineered the CytK nanopore (cytotoxin K) to be hydrophobic \emph{and} dipolar. CytK is a $\beta$-barrel pore-forming-toxin, homologous to the more established heptameric $\alpha$-haemolysin~\cite{sauciuc2024translocation,sauciuc2024blobs,versloot2022beta}. The wild type (WT) CytK does not exhibit self-gating at any voltage and its structure is robust enough to support multiple mutations~\cite{sauciuc2024translocation}.
We thus mutated both in experiments and simulations CytK pore to make its conducting barrel hydrophobic and to obtain a net dipole across the barrel; this was achieved using two rings of oppositely charged residues (Fig.~\ref{fig:cytk}a). 
A single negatively charged ring (Glu-139) was kept at the trans entrance, neutralizing the oppositely charged Lys-128, which was mutated into Met. To generate the dipolar environment surrounding the hydrophobic region, a positively charged ring was placed higher in the barrel towards the cis side, mutating Thr-147 into Lys. Finally, three hydrophilic residues (Thr-143, Ser-126, and Gln-145) were mutated into hydrophobic ones: we generated two hydrophobically gated variants, one with three Ile (M3IK) and one with two Ile and one Phe (M2IFK), see Fig.~\ref{fig:cytk}a.

The conductances of WT, M3IK, and M2IFK were measured in a 1M KCl solution at pH 7.5, at voltages ranging from -140 to +140 mV. Representative current traces are reported in Fig.~\ref{fig:cytk}b, showing that, after voltage application, the WT curve quickly reaches a stable value, while the hydrophobic mutants display the typical random telegraph behaviour of gating events. The mean conductances measured over many cycles are reported in Fig.~\ref{fig:cytk}c. In accordance with previous work~\cite{sauciuc2024translocation,sauciuc2024blobs,versloot2022beta}, the WT conductance is approximately constant with voltage with a value $0.85$~nS close to $\alpha$-hemolysin under similar conditions, corresponding to an heptameric nanopore.
The hydrophobically gated M3IK and M2FIK mutants display conductances that monotonically decrease with the applied voltage, suggesting that the mutated nanopores exhibit electrodrying over the whole voltage range that can be covered in experiments. 
Both mutants reach a (maximum open pore) conductance similar to that of the WT at -140 mV, underscoring that, despite its large steric hindrance, the introduction of the Phe-145 ring does not affect the wet-pore conductance, as also reported for similar mutants~\cite{versloot2022beta}.

For M3IK, the onset of electrodrying is most evident at positive voltages. In this range, the electric field aligns with the nominal dipole of the nanopore. This condition corresponds to the electrowetting regime based on the theory developed for pure water (Fig.~\ref{fig:system}c). However, experiments were performed in 1M KCl solution. To investigate the microscopic behaviour of the engineered nanopore in the presence of ions, the M3IK mutant was simulated with all-atoms MD under different applied voltages computing, for each voltage, the free-energy profiles of hydration (Figs.~\ref{fig:cytk}d,e). At $V=0$, MD simulations predict a bistable system with wet and dry states, with the wet state ($N_w=75$) being the most favourable ($\Delta F = -5\, \mathrm{k_B T}$). At $V = 150$~mV, which falls in the range of positive voltages explored in the experiments, the system displays electrodrying, in accordance with the experiments: the free energy of the wet state increases, indicating the destabilization of the hydrated pore. Instead, at negative voltages, the wet state is always more stable than for $V=0$. For $V =300$~mV, a value not attainable in experiments, the system falls again in the electrowetting regime, in agreement with the expectations of the theory and of the model nanopore (Figs.~\ref{fig:system}d and \ref{fig:MD-result}d).

The electrostatic potential calculated from equilibrium MD simulations of M3IK at 1\,M KCl is dominated by counterion accumulation close to the mutated Lys and Glu rings, leading to an unexpected charge inversion (Supplementary Fig.~S6). The resulting intrinsic potential drop along the channel axis is on average $V_\mathrm{int}= 140$~mV which has an opposite sign compared to that induced by the same charged amino-acid rings in vacuo. $V_\mathrm{int}$ sets the minimum conductance and the electrodrying range, as shown for the model nanopore in Fig.~\ref{fig:currents}a. Therefore, the unexpected observation of electrodrying in M3IK at positive voltages in both experiments and simulations can be explained by the dipole inversion observed for relatively concentrated electrolyte solutions.

MD predicts that the nanopore is fully wet and conductive for most of the time at $V = 0$ with a probability
$$
n_\mathrm{wet} = \frac{\tau_{wet}}{\tau_{wet} + \tau_{dry}} = \frac{1}{\exp(\Delta F/k_BT) + 1} \; ,
$$
which yields $n_\mathrm{wet}> 99\%$. 
Assuming that the open-pore conductance is independent of the voltage, an estimate of the conductivity can be calculated multiplying the open-pore conductance by the wet-pore probability computed from the simulations in Fig.~\ref{fig:cytk}d: $G=G_{o} n_\mathrm{wet} (V)$. $G_o$ is obtained by averaging the conductances at negative voltages. Figure~\ref{fig:cytk}c shows that both the value and the trend of the conductance estimated from simulations is in fair agreement with the experimental one.

One may wonder whether the presence of a dipole along the conducting channel could itself induce a certain degree of conductance rectification, even in the absence of gating events, i.e., $n_w=1$ and $G=G_o(V)$. In the cases reported in the literature~\cite{siwy2006ion,cervera2006ionic}, this kind of rectification is usually associated with pronounced geometrical asymmetry, highly charged environments, and/or the presence of intense electroosmotic flows~\cite{ramirez2008pore,laucirica2021nanofluidic,ai2010effects,baldelli2024performance}, which lead to significant variations in the open-pore conductance when the voltage is varied. 
In contrast, the present experiments show that the open-pore conductance of our engineered CytK nanopore decreases much less than the average conductance and, in several cases, remains comparable to that of the WT at voltages where the pore is predominantly closed (Fig.~\ref{fig:cytk}b and Supplementary Fig.~S5). Therefore, the changes in average conductance reported in Fig.~\ref{fig:cytk}c cannot be explained by voltage-induced variations in $G_o$ alone, but instead imply that the open-pore probability \emph{decreases} with increasing voltage, i.e., the reported conductance trend is consistent with electrodrying.

Together, experimental and simulation results on CytK demonstrate in a biotechnologically relevant nanopore that an external voltage can destabilise the wet state, in accordance with the proposed electrodrying concept and in contrast to the common understanding that voltage invariably promotes wetting.

\section{Conclusions}
This work introduces the concept of electrodrying, a counterintuitive physical phenomenon by which an external voltage reduces water occupancy in hydrophobic nanopores.

By integrating electrostatic theory, molecular dynamics simulations, and targeted experiments on engineered CytK nanopores, we establish a comprehensive framework for achieving bidirectional, dynamical tuning of nanopore conductance by precise electrostatic control. Among other applications, electrodrying open new avenues for realising iontronic and memristive devices.

Our work demonstrates that the balance between an intrinsic dipole across a nanopore and an applied voltage can be used to reliably switch the pore between wet (conductive) and dry (non-conductive) states. 
This bidirectional control over ion currents through nanopores significantly extends the scope of conventional electrowetting approaches to control conductance \cite{paulo2023hydrophobically}, which can only be used to switch on currents.
This mechanism was found to be robust across a range of dipole configurations and electrolyte concentrations, providing a broad design space for solid-state and biological systems.

Electrodrying nanopores display a peculiar electrical behaviour, characterized by asymmetric, non-monotonic, and hysteretic response to voltage cycles. On the one hand, the pinched hysteresis loop in the current-voltage curve demonstrates a memristive behaviour that was not previously reported \cite{sun2019}, with the unique possibility to offset the pinch point at will. On the other hand, the fact that electrodrying favours a non-conductive state implies a voltage range with negative differential resistance, a property that can be used in iontronic oscillators. Interestingly, NDR is also found in several voltage-gated ion channels, including sodium \cite{hille1970ionic} and inward rectifying potassium channels \cite{Tourneur1986}, further suggesting that electrodrying could be used to mimic the gating properties of ion channels, but without the need for complex structural rearrangements of biological counterparts.

In short, electrodrying emerges as a fascinating physical phenomenon that promises to impact our fundamental understanding of fluids in biological and synthetic nanopores and to propel novel neuromorphic and iontronic applications. 

\section{Methods}

\subsection*{Molecular Dynamics}

\textbf{Model pore setups.}
Following the same protocol of our previous work~\cite{Paulo2023jcp}, we built a molecular dynamics system for which we want to compute the free energy $F$ and the diffusivity $D$ associated with the number of water molecules $N$ inside a hydrophobic nanopore; $N$ represents the coarse-grained variable which defines the wetting/drying process.
The system, represented in Fig.~\ref{fig:system}, is made of a slab of fixed Lennard-Jones atoms in an fcc arrangement, with lattice spacing $0.35$~nm, from which a cylindrical nanopore is excavated. This slab is surrounded by water molecules (SPC/E~\cite{berendsen1987missing}), and the water-solid non-bonded interactions are  tuned~\cite{camisasca2020gas} so that the contact angle is ca. $104^\circ$. We also consider a system with Na and Cl atoms at 0.5M and 0.05M.

The nanopore has a diameter of 1.4~$nm$ and a length of 2.8~$nm$. Charge is added to a ring of atoms at the two pore mouths to create a dipole inside the nanopore while maintaining the electroneutrality of the system. The surface atoms are given partial electrical charges in such a way as to have the 3 total charges studied in this work, 0.5~$e$, 2.5~$e$, 5~$e$ and 10~$e$.
To control the pressure of the system, we used two pistons orthogonal to the pore axis~\cite{marchio2018}. The NVT ensemble was sampled using a Nos\'e--Hoover chains thermostat~\cite{martyna1992} at 310~K with a chain length of 3. A constant and homogeneous electric field is applied across the system $\mathrm{E}=(0,0,E_z)$, with $z$ being the direction parallel to the pore axis, to mimic a difference of a constant voltage across the membrane $V=E_z L_z$~\cite{gumbart2012constant}, with $L_z$ the length of the MD box. 

\textbf{CytK mutants Setup.}
The starting PDB structure of the CytK heptamer was obtained from a previous work~\cite{sauciuc2024controlled}. To introduce the engineered dipolar environment, key mutations were introduced into the wild-type structure (barrel shown in Fig.~\ref{fig:cytk}a, using the VMD Mutator Plugin~\cite{vmd}.
In particular, to establish a dipole across the pore, the residue Lys-128 at the trans entrance was mutated to Met, thereby neutralizing its negative charge, while Thr-147 at the cis-side barrel entrance was mutated to Lys to introduce a positive charge. In addition, the hydrophilic residues Thr-145, Ser-126, and Gln-145 were replaced with hydrophobic amino acids. Two variants were generated: one in which all three hydrophilic residues were substituted with isoleucine (M3IK) and another in which two were replaced by isoleucine and one by phenylalanine (M2IFK). 
The protonation states of all titratable residues were determined via PROPKA3~\cite{olsson2011propka3} at pH 7. The modified structures were minimized for 1000 steps of gradient descent in vacuum. Subsequently, the systems were embedded in a lipid bilayer and solvated in a 1M KCl solution using VMD’s Membrane Builder, Solvate, and Autoionize plugins. The final simulation system comprised approximately 260,000 atoms, including roughly 480 molecules of 1-palmitoyl-2-oleoyl-sn-glycero-3-phosphocholine (POPC), 50,000 water molecules, 1050 potassium ions, and 1000 chloride ions, contained within a simulation box of dimensions approximately 155 × 155 × 150~$\AA^3$. For the molecular dynamics simulations, the ff15ipq~\cite{debiec2016further} force field was applied to the protein, the Lipid17~\cite{dickson2014lipid14} force field to the phospholipids, and the SPC/E$_b$~\cite{takemura2012water} model was used for water. The systems were thermalized and equilibrated for 10 ns, following the protocol established in previous studies~\cite{baldelli2024controlling}.

\textbf{Free energy and diffusivity computation}
We use Restrained Molecular Dynamics (RMD)~\cite{maragliano2006temperature} to compute the free energy as a function of the pore filling $N$.
This is done by adding a harmonic restraint to the physical Hamiltonian $H_\mathrm{phys}$  of the system,
\begin{equation}\label{eq:restraint}
    H_N(\boldsymbol{r},\boldsymbol{p})=H_\mathrm{phys}(\boldsymbol{r},\boldsymbol{p})+\frac{k}{2}\left(N-\tilde{N}(\boldsymbol{r})\right)^2\,,
\end{equation}
where $\boldsymbol{r}$ and $\boldsymbol{p}$ are the positions and momenta of all the atoms, respectively, $k$ is a harmonic constant (1~kcal/mol), $N$ is the desired number of water molecules in a box centered around the nanopore, and $\tilde{N}$ is computed by counting the number of water molecules in the region highlighted in black in Fig.~\ref{fig:system}. 
The counting procedure has been explained in detail in previous work~\cite{Paulo2023jcp}.
The protocol is implemented in NAMD~\cite{phillips2005scalable} by using the Volumetric map-based variables of the Colvars Module~\cite{fiorin2013using}, as introduced in~\cite{fiorin2020direct}; see also~\cite{paulo2023hydrophobically,coronel2024lipid} for details on other nanopores and ion channel systems.

\subsection*{Experimental}

\textbf{Materials.}
The chemicals and suppliers used are listed as follows: Ampicillin sodium salt was purchased from Fisher Bio Reagents; chloramphenicol ($\geq$98.0) from Sigma Life Science), Isopropyl $\beta$-D-thiogalactopyranoside ($\geq$99.0\%, dioxin-free, animal-free), LB medium, NaCl ($\geq$99.5\%), HEPES (PUFFERAN® CELLPURE® ($\geq$99.5\%), imidazole ($\geq$99\%), KCl ($\geq$99.5\%), Tris(2-carboxyethyl)phosphine hydrochloride ($\geq$98.0\%), Dodecyl-$\beta$-D-maltoside ($\geq$99\%) from Roth; and n-hexadecane (99\%) from Acros Organics; protease inhibitors (Pierce™ Protease inhibitor Mini tablets, EDTA-free); GeneJET gel extraction kit, GeneJET PCR purification kit, GeneJET Plasmid Miniprep kit were purchased from (Thermo Scientific); Ni-NTA agarose from Qiagen; DPhPC from Avanti polar lipids, n-pentane from Sigma-Aldrich; Quick Start Bradford 1x Dye Reagent (Bio-Rad); DNA primers and gBlock™ from IDT. BL21(DE3) strain harbouring the pET-PfuX7 plasmid was kindly provided by prof. dr. Oscar Kuipers.

\textbf{Cloning.}
The mutants were generated using the plasmids encoding for CytK-WT and CytK-4D-S126I~\cite{sauciuc2024controlled}. 
PCRs and USER cloning were performed as previously described (\cite{sauciuc2024controlled,sauciuc2024translocation}). 
Proteins sequences and primers are reported in Supplementary Table~S1.
In order to introduce the mutations in one step comprising of two PCR rounds, the nanopore gene was split into two fragments: the upstream fragment was amplified from CytK-4D-S126I and the downstream fragment was amplified from CytK-WT. In the first PCR round, the upstream fragment was amplified with the T7 pro primer and a mutagenic primer containing the K128M mutation and an overhang (approximately corresponding to the A129-S140 bases). The two possible downstream fragments were obtained by amplification with the T7 term primer and a mutagenic primer annealing in the S134-A142 region and bearing an overhang with the T143I, Q145I/F and T147K mutations. The resulting 3 fragments were run on a 1.5\% agarose gel and gel extracted. In the second PCR round, the unique upstream fragment (containing the S126I and K128M mutations) was added to each downstream fragment (containing the T143I, Q145I/F and T147K mutations). The full genes encoding for the M3IK or M2IFK, respectively, mutant was obtained by elongating the two fragments with U-containing primers annealing upstream of the ATG start codon and downstream the stop codon. The full gene was introduced into an empty pT7-sc1 vector by means of USER cloning, as previously described. After plasmid amplification, the presence of mutations was confirmed by Sanger sequencing (Eurofins).

\textbf{Nanopore production.}
The plasmids (AmpR) containing the mutant nanopore genes were transformed into E. coli SoluBl21 (DE3) cells. The successful transformants were selected on LB-agar plates containing 100 ug/mL ampicillin and 1\% glucose. Several colonies were resuspended in LB medium containing 100 ug / ml of ampicillin so that the starting OD600 was 0.05-0.1. Cultures were grown at 37 C, 180 rpm until an OD600 value of 0.7-0.9, when cultures were chilled for a few min in water / ice and protein production was induced with 0.5 mM IPTG. Incubation was continued for 19-20 h at 25 C, 180 rpm. The cells were then harvested (7500 rpm, 6 min) and the resulting pellets were incubated at -70 C for at least 1 h. The mutant nanopores were purified using the His tag, as described in previous work (\cite{sauciuc2024translocation}). The presence of oligomers was assessed by running samples that were not preheated and the pores were stored at 4 C$^\circ$.

\textbf{Electrophysiology.}
Planar lipid bilayer recordings were performed using a chamber with two compartments separated by a 25 $\mu$m thick Teflon membrane with an aperture of approximately 100 $\mu$ m, as previously described\cite{maglia2010analysis}. To prepare the bilayer, a droplet containing 6.25\% n-hexadecane dissolved in n-pentane, equivalent to half the volume of a 10 $\mu$L glass capillary, was applied to the membrane. After the n-pentane evaporated, 500 $\mu$L of buffer solution (1 M KCl, 15 mM HEPES, pH 7.5) was introduced, followed by the addition of two to three droplets of DPhPC lipids (5 mg/mL in n-pentane) into each compartment. The Ag / AgCl electrodes were connected to both compartments through 2.5\% agarose bridges containing a 3 M KCl solution, with the trans compartment serving as the ground. Data acquisition was carried out using an Axon™ Digidata® 1550B digitizer and an Axopatch 200B amplifier (Molecular Devices), with recordings collected via Clampex 11.1 software.
The diluted (10, 50, 100 times in 150 mM NaCl, 50 mM HEPES, 0.02\% DDM pH 7.4) elution fractions from-tag purification were used in electrophysiology experiments. When a single insertion was observed, the nanopore was isolated by perfusing the cis chamber with 1 M KCl, 15 mM HEPES. Finally, the nanopore current was recorded at different applied potentials.

\section{Acknowledgement}

This research is part of a project that has received funding from the European Research Council (ERC) under the European Union's Horizon 2020 research and innovation programme (grant agreement No. 803213). 
We acknowledge financial support by the Italian Ministry of Education, University, and Research (MIUR) through the ``Framework per l'Attrazione e il Rafforzamento delle Eccellenze per la Ricerca in Italia (FARE)'' scheme, grant SERENA n. R18XYKRW7J.
We acknowledge EuroHPC Joint Undertaking for awarding us access to MareNostrum5 hosted at BSC, Spain. AS and GM acknowledge NWO-VICI VI.C.192.068 grant, and NHGRI 1R01HG012554. LI acknowledges financial support under the National Recovery and Resilience Plan (NRRP),
CUP B53C23002020006.

\section*{Data Availability Statement}
The data that support the findings of this study are openly available in Zenodo at DOI:
10.5281/zenodo.17417489

\bibliography{aipsamp}

\begin{thebibliography}{81}%
\makeatletter
\providecommand \@ifxundefined [1]{%
 \@ifx{#1\undefined}
}%
\providecommand \@ifnum [1]{%
 \ifnum #1\expandafter \@firstoftwo
 \else \expandafter \@secondoftwo
 \fi
}%
\providecommand \@ifx [1]{%
 \ifx #1\expandafter \@firstoftwo
 \else \expandafter \@secondoftwo
 \fi
}%
\providecommand \natexlab [1]{#1}%
\providecommand \enquote  [1]{``#1''}%
\providecommand \bibnamefont  [1]{#1}%
\providecommand \bibfnamefont [1]{#1}%
\providecommand \citenamefont [1]{#1}%
\providecommand \href@noop [0]{\@secondoftwo}%
\providecommand \href [0]{\begingroup \@sanitize@url \@href}%
\providecommand \@href[1]{\@@startlink{#1}\@@href}%
\providecommand \@@href[1]{\endgroup#1\@@endlink}%
\providecommand \@sanitize@url [0]{\catcode `\\12\catcode `\$12\catcode
  `\&12\catcode `\#12\catcode `\^12\catcode `\_12\catcode `\%12\relax}%
\providecommand \@@startlink[1]{}%
\providecommand \@@endlink[0]{}%
\providecommand \url  [0]{\begingroup\@sanitize@url \@url }%
\providecommand \@url [1]{\endgroup\@href {#1}{\urlprefix }}%
\providecommand \urlprefix  [0]{URL }%
\providecommand \Eprint [0]{\href }%
\providecommand \doibase [0]{http://dx.doi.org/}%
\providecommand \selectlanguage [0]{\@gobble}%
\providecommand \bibinfo  [0]{\@secondoftwo}%
\providecommand \bibfield  [0]{\@secondoftwo}%
\providecommand \translation [1]{[#1]}%
\providecommand \BibitemOpen [0]{}%
\providecommand \bibitemStop [0]{}%
\providecommand \bibitemNoStop [0]{.\EOS\space}%
\providecommand \EOS [0]{\spacefactor3000\relax}%
\providecommand \BibitemShut  [1]{\csname bibitem#1\endcsname}%
\let\auto@bib@innerbib\@empty
\bibitem [{\citenamefont {Lippmann}\ \emph {et~al.}(1875)\citenamefont
  {Lippmann} \emph {et~al.}}]{lippmann1875relations}%
  \BibitemOpen
  \bibfield  {author} {\bibinfo {author} {\bibfnamefont {G.}~\bibnamefont
  {Lippmann}} \emph {et~al.},\ }\href@noop {} {\bibfield  {journal} {\bibinfo
  {journal} {Ann. Chim. Phys}\ }\textbf {\bibinfo {volume} {5}},\ \bibinfo
  {pages} {494} (\bibinfo {year} {1875})}\BibitemShut {NoStop}%
\bibitem [{\citenamefont {Mugele}\ and\ \citenamefont
  {Baret}(2005)}]{Mugele2005}%
  \BibitemOpen
  \bibfield  {author} {\bibinfo {author} {\bibfnamefont {F.}~\bibnamefont
  {Mugele}}\ and\ \bibinfo {author} {\bibfnamefont {J.-C.}\ \bibnamefont
  {Baret}},\ }\href {\doibase 10.1088/0953-8984/17/28/r01} {\bibfield
  {journal} {\bibinfo  {journal} {Journal of Physics: Condensed Matter}\
  }\textbf {\bibinfo {volume} {17}},\ \bibinfo {pages} {R705} (\bibinfo {year}
  {2005})}\BibitemShut {NoStop}%
\bibitem [{\citenamefont {Jones}(2005)}]{jones2005electromechanical}%
  \BibitemOpen
  \bibfield  {author} {\bibinfo {author} {\bibfnamefont {T.~B.}\ \bibnamefont
  {Jones}},\ }\href@noop {} {\bibfield  {journal} {\bibinfo  {journal} {Journal
  of Micromechanics and Microengineering}\ }\textbf {\bibinfo {volume} {15}},\
  \bibinfo {pages} {1184} (\bibinfo {year} {2005})}\BibitemShut {NoStop}%
\bibitem [{\citenamefont {Lee}\ \emph {et~al.}(2002)\citenamefont {Lee},
  \citenamefont {Moon}, \citenamefont {Fowler}, \citenamefont {Schoellhammer},\
  and\ \citenamefont {Kim}}]{LEE2002259}%
  \BibitemOpen
  \bibfield  {author} {\bibinfo {author} {\bibfnamefont {J.}~\bibnamefont
  {Lee}}, \bibinfo {author} {\bibfnamefont {H.}~\bibnamefont {Moon}}, \bibinfo
  {author} {\bibfnamefont {J.}~\bibnamefont {Fowler}}, \bibinfo {author}
  {\bibfnamefont {T.}~\bibnamefont {Schoellhammer}}, \ and\ \bibinfo {author}
  {\bibfnamefont {C.-J.}\ \bibnamefont {Kim}},\ }\href {\doibase
  https://doi.org/10.1016/S0924-4247(01)00734-8} {\bibfield  {journal}
  {\bibinfo  {journal} {Sensors and Actuators A: Physical}\ }\textbf {\bibinfo
  {volume} {95}},\ \bibinfo {pages} {259} (\bibinfo {year} {2002})},\ \bibinfo
  {note} {papers from the Proceedings of the 14th IEEE Internat. Conf. on
  MicroElectroMechanical Systems}\BibitemShut {NoStop}%
\bibitem [{\citenamefont {Cooney}\ \emph {et~al.}(2006)\citenamefont {Cooney},
  \citenamefont {Chen}, \citenamefont {Emerling}, \citenamefont {Nadim},\ and\
  \citenamefont {Sterling}}]{cooney2006electrowetting}%
  \BibitemOpen
  \bibfield  {author} {\bibinfo {author} {\bibfnamefont {C.~G.}\ \bibnamefont
  {Cooney}}, \bibinfo {author} {\bibfnamefont {C.-Y.}\ \bibnamefont {Chen}},
  \bibinfo {author} {\bibfnamefont {M.~R.}\ \bibnamefont {Emerling}}, \bibinfo
  {author} {\bibfnamefont {A.}~\bibnamefont {Nadim}}, \ and\ \bibinfo {author}
  {\bibfnamefont {J.~D.}\ \bibnamefont {Sterling}},\ }\href@noop {} {\bibfield
  {journal} {\bibinfo  {journal} {Microfluidics and Nanofluidics}\ }\textbf
  {\bibinfo {volume} {2}},\ \bibinfo {pages} {435} (\bibinfo {year}
  {2006})}\BibitemShut {NoStop}%
\bibitem [{\citenamefont {Wikramanayake}\ \emph {et~al.}(2020)\citenamefont
  {Wikramanayake}, \citenamefont {Perry},\ and\ \citenamefont
  {Bahadur}}]{wikramanayake2020ac}%
  \BibitemOpen
  \bibfield  {author} {\bibinfo {author} {\bibfnamefont {E.~D.}\ \bibnamefont
  {Wikramanayake}}, \bibinfo {author} {\bibfnamefont {J.}~\bibnamefont
  {Perry}}, \ and\ \bibinfo {author} {\bibfnamefont {V.}~\bibnamefont
  {Bahadur}},\ }\href@noop {} {\bibfield  {journal} {\bibinfo  {journal}
  {Applied Physics Letters}\ }\textbf {\bibinfo {volume} {116}} (\bibinfo
  {year} {2020})}\BibitemShut {NoStop}%
\bibitem [{\citenamefont {Krupenkin}\ and\ \citenamefont
  {Taylor}(2011)}]{krupenkin2011}%
  \BibitemOpen
  \bibfield  {author} {\bibinfo {author} {\bibfnamefont {T.}~\bibnamefont
  {Krupenkin}}\ and\ \bibinfo {author} {\bibfnamefont {J.~A.}\ \bibnamefont
  {Taylor}},\ }\href@noop {} {\bibfield  {journal} {\bibinfo  {journal} {Nature
  communications}\ }\textbf {\bibinfo {volume} {2}},\ \bibinfo {pages} {448}
  (\bibinfo {year} {2011})}\BibitemShut {NoStop}%
\bibitem [{\citenamefont {Boroujeni}\ \emph {et~al.}(2020)\citenamefont
  {Boroujeni}, \citenamefont {Raissi}, \citenamefont {Jafarabadi-Ashtiani},
  \citenamefont {Riahifar},\ and\ \citenamefont
  {Sahba-Yaghmaee}}]{boroujeni2020droplet}%
  \BibitemOpen
  \bibfield  {author} {\bibinfo {author} {\bibfnamefont {F.~G.}\ \bibnamefont
  {Boroujeni}}, \bibinfo {author} {\bibfnamefont {B.}~\bibnamefont {Raissi}},
  \bibinfo {author} {\bibfnamefont {S.}~\bibnamefont {Jafarabadi-Ashtiani}},
  \bibinfo {author} {\bibfnamefont {R.}~\bibnamefont {Riahifar}}, \ and\
  \bibinfo {author} {\bibfnamefont {M.}~\bibnamefont {Sahba-Yaghmaee}},\
  }\href@noop {} {\bibfield  {journal} {\bibinfo  {journal} {Engineering
  Research Express}\ }\textbf {\bibinfo {volume} {2}},\ \bibinfo {pages}
  {045028} (\bibinfo {year} {2020})}\BibitemShut {NoStop}%
\bibitem [{\citenamefont {Riaud}\ \emph {et~al.}(2021)\citenamefont {Riaud},
  \citenamefont {Wang}, \citenamefont {Zhou}, \citenamefont {Xu},\ and\
  \citenamefont {Wang}}]{riaud2021hydrodynamic}%
  \BibitemOpen
  \bibfield  {author} {\bibinfo {author} {\bibfnamefont {A.}~\bibnamefont
  {Riaud}}, \bibinfo {author} {\bibfnamefont {C.}~\bibnamefont {Wang}},
  \bibinfo {author} {\bibfnamefont {J.}~\bibnamefont {Zhou}}, \bibinfo {author}
  {\bibfnamefont {W.}~\bibnamefont {Xu}}, \ and\ \bibinfo {author}
  {\bibfnamefont {Z.}~\bibnamefont {Wang}},\ }\href@noop {} {\bibfield
  {journal} {\bibinfo  {journal} {Microsystems \& Nanoengineering}\ }\textbf
  {\bibinfo {volume} {7}},\ \bibinfo {pages} {49} (\bibinfo {year}
  {2021})}\BibitemShut {NoStop}%
\bibitem [{\citenamefont {Grahame}(1947)}]{Grahame1947}%
  \BibitemOpen
  \bibfield  {author} {\bibinfo {author} {\bibfnamefont {D.~C.}\ \bibnamefont
  {Grahame}},\ }\href {\doibase 10.1021/cr60130a002} {\bibfield  {journal}
  {\bibinfo  {journal} {Chemical Reviews}\ }\textbf {\bibinfo {volume} {41}},\
  \bibinfo {pages} {441–501} (\bibinfo {year} {1947})}\BibitemShut {NoStop}%
\bibitem [{\citenamefont {Zeng}\ and\ \citenamefont
  {Korsmeyer}(2004)}]{Zeng2004}%
  \BibitemOpen
  \bibfield  {author} {\bibinfo {author} {\bibfnamefont {J.}~\bibnamefont
  {Zeng}}\ and\ \bibinfo {author} {\bibfnamefont {T.}~\bibnamefont
  {Korsmeyer}},\ }\href {\doibase 10.1039/b403082f} {\bibfield  {journal}
  {\bibinfo  {journal} {Lab on a Chip}\ }\textbf {\bibinfo {volume} {4}},\
  \bibinfo {pages} {265} (\bibinfo {year} {2004})}\BibitemShut {NoStop}%
\bibitem [{\citenamefont {Teh}\ \emph {et~al.}(2008)\citenamefont {Teh},
  \citenamefont {Lin}, \citenamefont {Hung},\ and\ \citenamefont
  {Lee}}]{Teh2008}%
  \BibitemOpen
  \bibfield  {author} {\bibinfo {author} {\bibfnamefont {S.-Y.}\ \bibnamefont
  {Teh}}, \bibinfo {author} {\bibfnamefont {R.}~\bibnamefont {Lin}}, \bibinfo
  {author} {\bibfnamefont {L.-H.}\ \bibnamefont {Hung}}, \ and\ \bibinfo
  {author} {\bibfnamefont {A.~P.}\ \bibnamefont {Lee}},\ }\href {\doibase
  10.1039/b715524g} {\bibfield  {journal} {\bibinfo  {journal} {Lab on a Chip}\
  }\textbf {\bibinfo {volume} {8}},\ \bibinfo {pages} {198} (\bibinfo {year}
  {2008})}\BibitemShut {NoStop}%
\bibitem [{\citenamefont {Nelson}\ and\ \citenamefont
  {Kim}(2012)}]{Nelson2012}%
  \BibitemOpen
  \bibfield  {author} {\bibinfo {author} {\bibfnamefont {W.~C.}\ \bibnamefont
  {Nelson}}\ and\ \bibinfo {author} {\bibfnamefont {C.-J.}\ \bibnamefont
  {Kim}},\ }\href {\doibase 10.1163/156856111x599562} {\bibfield  {journal}
  {\bibinfo  {journal} {Journal of Adhesion Science and Technology}\ }\textbf
  {\bibinfo {volume} {26}},\ \bibinfo {pages} {1747–1771} (\bibinfo {year}
  {2012})}\BibitemShut {NoStop}%
\bibitem [{\citenamefont {Wu}\ \emph {et~al.}(2024)\citenamefont {Wu},
  \citenamefont {Li}, \citenamefont {Lin}, \citenamefont {Zhuang},
  \citenamefont {Tang}, \citenamefont {Liu},\ and\ \citenamefont
  {Zhou}}]{Wu2024}%
  \BibitemOpen
  \bibfield  {author} {\bibinfo {author} {\bibfnamefont {J.}~\bibnamefont
  {Wu}}, \bibinfo {author} {\bibfnamefont {X.}~\bibnamefont {Li}}, \bibinfo
  {author} {\bibfnamefont {T.}~\bibnamefont {Lin}}, \bibinfo {author}
  {\bibfnamefont {L.}~\bibnamefont {Zhuang}}, \bibinfo {author} {\bibfnamefont
  {B.}~\bibnamefont {Tang}}, \bibinfo {author} {\bibfnamefont {F.}~\bibnamefont
  {Liu}}, \ and\ \bibinfo {author} {\bibfnamefont {G.}~\bibnamefont {Zhou}},\
  }\href {\doibase 10.1021/acsami.3c13792} {\bibfield  {journal} {\bibinfo
  {journal} {ACS Applied Materials \& Interfaces}\ } (\bibinfo {year} {2024}),\
  10.1021/acsami.3c13792}\BibitemShut {NoStop}%
\bibitem [{\citenamefont {Dzubiella}\ \emph {et~al.}(2004)\citenamefont
  {Dzubiella}, \citenamefont {Allen},\ and\ \citenamefont
  {Hansen}}]{Dzubiella2004}%
  \BibitemOpen
  \bibfield  {author} {\bibinfo {author} {\bibfnamefont {J.}~\bibnamefont
  {Dzubiella}}, \bibinfo {author} {\bibfnamefont {R.}~\bibnamefont {Allen}}, \
  and\ \bibinfo {author} {\bibfnamefont {J.-P.}\ \bibnamefont {Hansen}},\
  }\href@noop {} {\bibfield  {journal} {\bibinfo  {journal} {The Journal of
  chemical physics}\ }\textbf {\bibinfo {volume} {120}},\ \bibinfo {pages}
  {5001} (\bibinfo {year} {2004})}\BibitemShut {NoStop}%
\bibitem [{\citenamefont {Dzubiella}\ and\ \citenamefont
  {Hansen}(2005)}]{Dzubiella2005}%
  \BibitemOpen
  \bibfield  {author} {\bibinfo {author} {\bibfnamefont {J.}~\bibnamefont
  {Dzubiella}}\ and\ \bibinfo {author} {\bibfnamefont {J.-P.}\ \bibnamefont
  {Hansen}},\ }\href {\doibase 10.1063/1.1927514} {\bibfield  {journal}
  {\bibinfo  {journal} {The Journal of Chemical Physics}\ }\textbf {\bibinfo
  {volume} {122}} (\bibinfo {year} {2005}),\ 10.1063/1.1927514}\BibitemShut
  {NoStop}%
\bibitem [{\citenamefont {Powell}\ \emph {et~al.}(2011)\citenamefont {Powell},
  \citenamefont {Cleary}, \citenamefont {Davenport}, \citenamefont {Shea},\
  and\ \citenamefont {Siwy}}]{Powell2011}%
  \BibitemOpen
  \bibfield  {author} {\bibinfo {author} {\bibfnamefont {M.~R.}\ \bibnamefont
  {Powell}}, \bibinfo {author} {\bibfnamefont {L.}~\bibnamefont {Cleary}},
  \bibinfo {author} {\bibfnamefont {M.}~\bibnamefont {Davenport}}, \bibinfo
  {author} {\bibfnamefont {K.~J.}\ \bibnamefont {Shea}}, \ and\ \bibinfo
  {author} {\bibfnamefont {Z.~S.}\ \bibnamefont {Siwy}},\ }\href {\doibase
  10.1038/nnano.2011.189} {\bibfield  {journal} {\bibinfo  {journal} {Nature
  Nanotechnology}\ }\textbf {\bibinfo {volume} {6}},\ \bibinfo {pages} {798}
  (\bibinfo {year} {2011})}\BibitemShut {NoStop}%
\bibitem [{\citenamefont {Smirnov}\ \emph {et~al.}(2011)\citenamefont
  {Smirnov}, \citenamefont {Vlassiouk},\ and\ \citenamefont
  {Lavrik}}]{Smirnov2011}%
  \BibitemOpen
  \bibfield  {author} {\bibinfo {author} {\bibfnamefont {S.~N.}\ \bibnamefont
  {Smirnov}}, \bibinfo {author} {\bibfnamefont {I.~V.}\ \bibnamefont
  {Vlassiouk}}, \ and\ \bibinfo {author} {\bibfnamefont {N.~V.}\ \bibnamefont
  {Lavrik}},\ }\href {\doibase 10.1021/nn202392d} {\bibfield  {journal}
  {\bibinfo  {journal} {{ACS} Nano}\ }\textbf {\bibinfo {volume} {5}},\
  \bibinfo {pages} {7453} (\bibinfo {year} {2011})}\BibitemShut {NoStop}%
\bibitem [{\citenamefont {Roth}\ \emph {et~al.}(2008)\citenamefont {Roth},
  \citenamefont {Gillespie}, \citenamefont {Nonner},\ and\ \citenamefont
  {Eisenberg}}]{roth2008bubbles}%
  \BibitemOpen
  \bibfield  {author} {\bibinfo {author} {\bibfnamefont {R.}~\bibnamefont
  {Roth}}, \bibinfo {author} {\bibfnamefont {D.}~\bibnamefont {Gillespie}},
  \bibinfo {author} {\bibfnamefont {W.}~\bibnamefont {Nonner}}, \ and\ \bibinfo
  {author} {\bibfnamefont {R.~E.}\ \bibnamefont {Eisenberg}},\ }\href@noop {}
  {\bibfield  {journal} {\bibinfo  {journal} {Biophysical journal}\ }\textbf
  {\bibinfo {volume} {94}},\ \bibinfo {pages} {4282} (\bibinfo {year}
  {2008})}\BibitemShut {NoStop}%
\bibitem [{\citenamefont {Aryal}\ \emph {et~al.}(2015)\citenamefont {Aryal},
  \citenamefont {Sansom},\ and\ \citenamefont {Tucker}}]{aryal2015}%
  \BibitemOpen
  \bibfield  {author} {\bibinfo {author} {\bibfnamefont {P.}~\bibnamefont
  {Aryal}}, \bibinfo {author} {\bibfnamefont {M.~S.}\ \bibnamefont {Sansom}}, \
  and\ \bibinfo {author} {\bibfnamefont {S.~J.}\ \bibnamefont {Tucker}},\
  }\href@noop {} {\bibfield  {journal} {\bibinfo  {journal} {Journal of
  molecular biology}\ }\textbf {\bibinfo {volume} {427}},\ \bibinfo {pages}
  {121} (\bibinfo {year} {2015})}\BibitemShut {NoStop}%
\bibitem [{\citenamefont {Giacomello}\ and\ \citenamefont
  {Roth}(2020)}]{giacomello2020bubble}%
  \BibitemOpen
  \bibfield  {author} {\bibinfo {author} {\bibfnamefont {A.}~\bibnamefont
  {Giacomello}}\ and\ \bibinfo {author} {\bibfnamefont {R.}~\bibnamefont
  {Roth}},\ }\href@noop {} {\bibfield  {journal} {\bibinfo  {journal} {Advances
  in Physics: X}\ }\textbf {\bibinfo {volume} {5}},\ \bibinfo {pages} {1817780}
  (\bibinfo {year} {2020})}\BibitemShut {NoStop}%
\bibitem [{\citenamefont {Zhu}\ and\ \citenamefont
  {Hummer}(2012)}]{zhu2012theory}%
  \BibitemOpen
  \bibfield  {author} {\bibinfo {author} {\bibfnamefont {F.}~\bibnamefont
  {Zhu}}\ and\ \bibinfo {author} {\bibfnamefont {G.}~\bibnamefont {Hummer}},\
  }\href@noop {} {\bibfield  {journal} {\bibinfo  {journal} {Journal of
  chemical theory and computation}\ }\textbf {\bibinfo {volume} {8}},\ \bibinfo
  {pages} {3759} (\bibinfo {year} {2012})}\BibitemShut {NoStop}%
\bibitem [{\citenamefont {Innes}\ \emph {et~al.}(2015)\citenamefont {Innes},
  \citenamefont {Gutierrez}, \citenamefont {Mann}, \citenamefont {Buchsbaum},\
  and\ \citenamefont {Siwy}}]{Innes2015}%
  \BibitemOpen
  \bibfield  {author} {\bibinfo {author} {\bibfnamefont {L.}~\bibnamefont
  {Innes}}, \bibinfo {author} {\bibfnamefont {D.}~\bibnamefont {Gutierrez}},
  \bibinfo {author} {\bibfnamefont {W.}~\bibnamefont {Mann}}, \bibinfo {author}
  {\bibfnamefont {S.~F.}\ \bibnamefont {Buchsbaum}}, \ and\ \bibinfo {author}
  {\bibfnamefont {Z.~S.}\ \bibnamefont {Siwy}},\ }\href {\doibase
  10.1039/c4an02244k} {\bibfield  {journal} {\bibinfo  {journal} {The Analyst}\
  }\textbf {\bibinfo {volume} {140}},\ \bibinfo {pages} {4804–4812} (\bibinfo
  {year} {2015})}\BibitemShut {NoStop}%
\bibitem [{\citenamefont {Xiao}\ \emph {et~al.}(2016)\citenamefont {Xiao},
  \citenamefont {Zhou}, \citenamefont {Kong}, \citenamefont {Xie},
  \citenamefont {Li}, \citenamefont {Zhang}, \citenamefont {Wen},\ and\
  \citenamefont {Jiang}}]{Xiao2016}%
  \BibitemOpen
  \bibfield  {author} {\bibinfo {author} {\bibfnamefont {K.}~\bibnamefont
  {Xiao}}, \bibinfo {author} {\bibfnamefont {Y.}~\bibnamefont {Zhou}}, \bibinfo
  {author} {\bibfnamefont {X.-Y.}\ \bibnamefont {Kong}}, \bibinfo {author}
  {\bibfnamefont {G.}~\bibnamefont {Xie}}, \bibinfo {author} {\bibfnamefont
  {P.}~\bibnamefont {Li}}, \bibinfo {author} {\bibfnamefont {Z.}~\bibnamefont
  {Zhang}}, \bibinfo {author} {\bibfnamefont {L.}~\bibnamefont {Wen}}, \ and\
  \bibinfo {author} {\bibfnamefont {L.}~\bibnamefont {Jiang}},\ }\href
  {\doibase 10.1021/acsnano.6b05682} {\bibfield  {journal} {\bibinfo  {journal}
  {ACS Nano}\ }\textbf {\bibinfo {volume} {10}},\ \bibinfo {pages}
  {9703–9709} (\bibinfo {year} {2016})}\BibitemShut {NoStop}%
\bibitem [{\citenamefont {Polster}\ \emph {et~al.}(2022)\citenamefont
  {Polster}, \citenamefont {Aydin}, \citenamefont {de~Souza}, \citenamefont
  {Bazant}, \citenamefont {Pham},\ and\ \citenamefont {Siwy}}]{Polster2022}%
  \BibitemOpen
  \bibfield  {author} {\bibinfo {author} {\bibfnamefont {J.~W.}\ \bibnamefont
  {Polster}}, \bibinfo {author} {\bibfnamefont {F.}~\bibnamefont {Aydin}},
  \bibinfo {author} {\bibfnamefont {J.~P.}\ \bibnamefont {de~Souza}}, \bibinfo
  {author} {\bibfnamefont {M.~Z.}\ \bibnamefont {Bazant}}, \bibinfo {author}
  {\bibfnamefont {T.~A.}\ \bibnamefont {Pham}}, \ and\ \bibinfo {author}
  {\bibfnamefont {Z.~S.}\ \bibnamefont {Siwy}},\ }\href {\doibase
  10.1021/jacs.2c03436} {\bibfield  {journal} {\bibinfo  {journal} {Journal of
  the American Chemical Society}\ }\textbf {\bibinfo {volume} {144}},\ \bibinfo
  {pages} {11693} (\bibinfo {year} {2022})}\BibitemShut {NoStop}%
\bibitem [{\citenamefont {Polster}\ \emph {et~al.}(2020)\citenamefont
  {Polster}, \citenamefont {Acar}, \citenamefont {Aydin}, \citenamefont {Zhan},
  \citenamefont {Pham},\ and\ \citenamefont {Siwy}}]{Polster2020}%
  \BibitemOpen
  \bibfield  {author} {\bibinfo {author} {\bibfnamefont {J.~W.}\ \bibnamefont
  {Polster}}, \bibinfo {author} {\bibfnamefont {E.~T.}\ \bibnamefont {Acar}},
  \bibinfo {author} {\bibfnamefont {F.}~\bibnamefont {Aydin}}, \bibinfo
  {author} {\bibfnamefont {C.}~\bibnamefont {Zhan}}, \bibinfo {author}
  {\bibfnamefont {T.~A.}\ \bibnamefont {Pham}}, \ and\ \bibinfo {author}
  {\bibfnamefont {Z.~S.}\ \bibnamefont {Siwy}},\ }\href {\doibase
  10.1021/acsnano.9b09777} {\bibfield  {journal} {\bibinfo  {journal} {ACS
  Nano}\ }\textbf {\bibinfo {volume} {14}},\ \bibinfo {pages} {4306–4315}
  (\bibinfo {year} {2020})}\BibitemShut {NoStop}%
\bibitem [{\citenamefont {Xu}\ \emph {et~al.}(2011)\citenamefont {Xu},
  \citenamefont {Qiao}, \citenamefont {Zhou},\ and\ \citenamefont
  {Chen}}]{Xu2011}%
  \BibitemOpen
  \bibfield  {author} {\bibinfo {author} {\bibfnamefont {B.}~\bibnamefont
  {Xu}}, \bibinfo {author} {\bibfnamefont {Y.}~\bibnamefont {Qiao}}, \bibinfo
  {author} {\bibfnamefont {Q.}~\bibnamefont {Zhou}}, \ and\ \bibinfo {author}
  {\bibfnamefont {X.}~\bibnamefont {Chen}},\ }\href {\doibase
  10.1021/la200477y} {\bibfield  {journal} {\bibinfo  {journal} {Langmuir}\
  }\textbf {\bibinfo {volume} {27}},\ \bibinfo {pages} {6349–6357} (\bibinfo
  {year} {2011})}\BibitemShut {NoStop}%
\bibitem [{\citenamefont {Gao}\ \emph {et~al.}(2022)\citenamefont {Gao},
  \citenamefont {Yin}, \citenamefont {Zhang},\ and\ \citenamefont
  {Xu}}]{Gao2022}%
  \BibitemOpen
  \bibfield  {author} {\bibinfo {author} {\bibfnamefont {Y.}~\bibnamefont
  {Gao}}, \bibinfo {author} {\bibfnamefont {M.}~\bibnamefont {Yin}}, \bibinfo
  {author} {\bibfnamefont {H.}~\bibnamefont {Zhang}}, \ and\ \bibinfo {author}
  {\bibfnamefont {B.}~\bibnamefont {Xu}},\ }\href {\doibase
  10.1021/acsnano.2c02240} {\bibfield  {journal} {\bibinfo  {journal} {ACS
  Nano}\ }\textbf {\bibinfo {volume} {16}},\ \bibinfo {pages} {9420–9427}
  (\bibinfo {year} {2022})}\BibitemShut {NoStop}%
\bibitem [{\citenamefont {Vanzo}\ \emph {et~al.}(2014)\citenamefont {Vanzo},
  \citenamefont {Bratko},\ and\ \citenamefont {Luzar}}]{Vanzo2014}%
  \BibitemOpen
  \bibfield  {author} {\bibinfo {author} {\bibfnamefont {D.}~\bibnamefont
  {Vanzo}}, \bibinfo {author} {\bibfnamefont {D.}~\bibnamefont {Bratko}}, \
  and\ \bibinfo {author} {\bibfnamefont {A.}~\bibnamefont {Luzar}},\ }\href
  {\doibase 10.1021/jp506389p} {\bibfield  {journal} {\bibinfo  {journal} {The
  Journal of Physical Chemistry B}\ }\textbf {\bibinfo {volume} {119}},\
  \bibinfo {pages} {8890} (\bibinfo {year} {2014})}\BibitemShut {NoStop}%
\bibitem [{\citenamefont {Pireddu}\ and\ \citenamefont
  {Rotenberg}(2023)}]{rotenberg2023}%
  \BibitemOpen
  \bibfield  {author} {\bibinfo {author} {\bibfnamefont {G.}~\bibnamefont
  {Pireddu}}\ and\ \bibinfo {author} {\bibfnamefont {B.}~\bibnamefont
  {Rotenberg}},\ }\href {\doibase 10.1103/PhysRevLett.130.098001} {\bibfield
  {journal} {\bibinfo  {journal} {Phys. Rev. Lett.}\ }\textbf {\bibinfo
  {volume} {130}},\ \bibinfo {pages} {098001} (\bibinfo {year}
  {2023})}\BibitemShut {NoStop}%
\bibitem [{\citenamefont {Najem}\ \emph {et~al.}(2018)\citenamefont {Najem},
  \citenamefont {Taylor}, \citenamefont {Weiss}, \citenamefont {Hasan},
  \citenamefont {Rose}, \citenamefont {Schuman}, \citenamefont {Belianinov},
  \citenamefont {Collier},\ and\ \citenamefont {Sarles}}]{Najem2018}%
  \BibitemOpen
  \bibfield  {author} {\bibinfo {author} {\bibfnamefont {J.~S.}\ \bibnamefont
  {Najem}}, \bibinfo {author} {\bibfnamefont {G.~J.}\ \bibnamefont {Taylor}},
  \bibinfo {author} {\bibfnamefont {R.~J.}\ \bibnamefont {Weiss}}, \bibinfo
  {author} {\bibfnamefont {M.~S.}\ \bibnamefont {Hasan}}, \bibinfo {author}
  {\bibfnamefont {G.}~\bibnamefont {Rose}}, \bibinfo {author} {\bibfnamefont
  {C.~D.}\ \bibnamefont {Schuman}}, \bibinfo {author} {\bibfnamefont
  {A.}~\bibnamefont {Belianinov}}, \bibinfo {author} {\bibfnamefont {C.~P.}\
  \bibnamefont {Collier}}, \ and\ \bibinfo {author} {\bibfnamefont {S.~A.}\
  \bibnamefont {Sarles}},\ }\href {\doibase 10.1021/acsnano.8b01282} {\bibfield
   {journal} {\bibinfo  {journal} {{ACS} Nano}\ }\textbf {\bibinfo {volume}
  {12}},\ \bibinfo {pages} {4702} (\bibinfo {year} {2018})}\BibitemShut
  {NoStop}%
\bibitem [{\citenamefont {Robin}\ \emph {et~al.}(2021)\citenamefont {Robin},
  \citenamefont {Kavokine},\ and\ \citenamefont {Bocquet}}]{robin2021modeling}%
  \BibitemOpen
  \bibfield  {author} {\bibinfo {author} {\bibfnamefont {P.}~\bibnamefont
  {Robin}}, \bibinfo {author} {\bibfnamefont {N.}~\bibnamefont {Kavokine}}, \
  and\ \bibinfo {author} {\bibfnamefont {L.}~\bibnamefont {Bocquet}},\
  }\href@noop {} {\bibfield  {journal} {\bibinfo  {journal} {Science}\ }\textbf
  {\bibinfo {volume} {373}},\ \bibinfo {pages} {687} (\bibinfo {year}
  {2021})}\BibitemShut {NoStop}%
\bibitem [{\citenamefont {Xiong}\ \emph {et~al.}(2023)\citenamefont {Xiong},
  \citenamefont {Li}, \citenamefont {He}, \citenamefont {Xie}, \citenamefont
  {Zong}, \citenamefont {Jiang}, \citenamefont {Ma}, \citenamefont {Wu},
  \citenamefont {Fei}, \citenamefont {Yu},\ and\ \citenamefont
  {Mao}}]{Xiong2023}%
  \BibitemOpen
  \bibfield  {author} {\bibinfo {author} {\bibfnamefont {T.}~\bibnamefont
  {Xiong}}, \bibinfo {author} {\bibfnamefont {C.}~\bibnamefont {Li}}, \bibinfo
  {author} {\bibfnamefont {X.}~\bibnamefont {He}}, \bibinfo {author}
  {\bibfnamefont {B.}~\bibnamefont {Xie}}, \bibinfo {author} {\bibfnamefont
  {J.}~\bibnamefont {Zong}}, \bibinfo {author} {\bibfnamefont {Y.}~\bibnamefont
  {Jiang}}, \bibinfo {author} {\bibfnamefont {W.}~\bibnamefont {Ma}}, \bibinfo
  {author} {\bibfnamefont {F.}~\bibnamefont {Wu}}, \bibinfo {author}
  {\bibfnamefont {J.}~\bibnamefont {Fei}}, \bibinfo {author} {\bibfnamefont
  {P.}~\bibnamefont {Yu}}, \ and\ \bibinfo {author} {\bibfnamefont
  {L.}~\bibnamefont {Mao}},\ }\href {\doibase 10.1126/science.adc9150}
  {\bibfield  {journal} {\bibinfo  {journal} {Science}\ }\textbf {\bibinfo
  {volume} {379}},\ \bibinfo {pages} {156} (\bibinfo {year}
  {2023})}\BibitemShut {NoStop}%
\bibitem [{\citenamefont {Ramirez}\ \emph {et~al.}(2023)\citenamefont
  {Ramirez}, \citenamefont {G{\'o}mez}, \citenamefont {Cervera}, \citenamefont
  {Mafe},\ and\ \citenamefont {Bisquert}}]{ramirez2023synaptical}%
  \BibitemOpen
  \bibfield  {author} {\bibinfo {author} {\bibfnamefont {P.}~\bibnamefont
  {Ramirez}}, \bibinfo {author} {\bibfnamefont {V.}~\bibnamefont {G{\'o}mez}},
  \bibinfo {author} {\bibfnamefont {J.}~\bibnamefont {Cervera}}, \bibinfo
  {author} {\bibfnamefont {S.}~\bibnamefont {Mafe}}, \ and\ \bibinfo {author}
  {\bibfnamefont {J.}~\bibnamefont {Bisquert}},\ }\href@noop {} {\bibfield
  {journal} {\bibinfo  {journal} {The Journal of Physical Chemistry Letters}\
  }\textbf {\bibinfo {volume} {14}},\ \bibinfo {pages} {10930} (\bibinfo {year}
  {2023})}\BibitemShut {NoStop}%
\bibitem [{\citenamefont {Paulo}\ \emph
  {et~al.}(2023{\natexlab{a}})\citenamefont {Paulo}, \citenamefont {Sun},
  \citenamefont {Di~Muccio}, \citenamefont {Gubbiotti}, \citenamefont
  {Morozzo~della Rocca}, \citenamefont {Geng}, \citenamefont {Maglia},
  \citenamefont {Chinappi},\ and\ \citenamefont
  {Giacomello}}]{paulo2023hydrophobically}%
  \BibitemOpen
  \bibfield  {author} {\bibinfo {author} {\bibfnamefont {G.}~\bibnamefont
  {Paulo}}, \bibinfo {author} {\bibfnamefont {K.}~\bibnamefont {Sun}}, \bibinfo
  {author} {\bibfnamefont {G.}~\bibnamefont {Di~Muccio}}, \bibinfo {author}
  {\bibfnamefont {A.}~\bibnamefont {Gubbiotti}}, \bibinfo {author}
  {\bibfnamefont {B.}~\bibnamefont {Morozzo~della Rocca}}, \bibinfo {author}
  {\bibfnamefont {J.}~\bibnamefont {Geng}}, \bibinfo {author} {\bibfnamefont
  {G.}~\bibnamefont {Maglia}}, \bibinfo {author} {\bibfnamefont
  {M.}~\bibnamefont {Chinappi}}, \ and\ \bibinfo {author} {\bibfnamefont
  {A.}~\bibnamefont {Giacomello}},\ }\href@noop {} {\bibfield  {journal}
  {\bibinfo  {journal} {Nature Communications}\ }\textbf {\bibinfo {volume}
  {14}},\ \bibinfo {pages} {8390} (\bibinfo {year}
  {2023}{\natexlab{a}})}\BibitemShut {NoStop}%
\bibitem [{\citenamefont {Klesse}\ \emph {et~al.}(2020)\citenamefont {Klesse},
  \citenamefont {Tucker},\ and\ \citenamefont {Sansom}}]{klesse2020}%
  \BibitemOpen
  \bibfield  {author} {\bibinfo {author} {\bibfnamefont {G.}~\bibnamefont
  {Klesse}}, \bibinfo {author} {\bibfnamefont {S.~J.}\ \bibnamefont {Tucker}},
  \ and\ \bibinfo {author} {\bibfnamefont {M.~S.~P.}\ \bibnamefont {Sansom}},\
  }\href {\doibase 10.1021/acsnano.0c04387} {\bibfield  {journal} {\bibinfo
  {journal} {{ACS} Nano}\ }\textbf {\bibinfo {volume} {14}},\ \bibinfo {pages}
  {10480} (\bibinfo {year} {2020})}\BibitemShut {NoStop}%
\bibitem [{\citenamefont {Kayal}\ and\ \citenamefont
  {Chandra}(2015)}]{Kayal2015}%
  \BibitemOpen
  \bibfield  {author} {\bibinfo {author} {\bibfnamefont {A.}~\bibnamefont
  {Kayal}}\ and\ \bibinfo {author} {\bibfnamefont {A.}~\bibnamefont
  {Chandra}},\ }\href {\doibase 10.1063/1.4936939} {\bibfield  {journal}
  {\bibinfo  {journal} {The Journal of Chemical Physics}\ }\textbf {\bibinfo
  {volume} {143}} (\bibinfo {year} {2015}),\ 10.1063/1.4936939}\BibitemShut
  {NoStop}%
\bibitem [{\citenamefont {Paulo}\ \emph {et~al.}(2024)\citenamefont {Paulo},
  \citenamefont {Gubbiotti}, \citenamefont {Di~Muccio},\ and\ \citenamefont
  {Giacomello}}]{paulo2024voltage}%
  \BibitemOpen
  \bibfield  {author} {\bibinfo {author} {\bibfnamefont {G.}~\bibnamefont
  {Paulo}}, \bibinfo {author} {\bibfnamefont {A.}~\bibnamefont {Gubbiotti}},
  \bibinfo {author} {\bibfnamefont {G.}~\bibnamefont {Di~Muccio}}, \ and\
  \bibinfo {author} {\bibfnamefont {A.}~\bibnamefont {Giacomello}},\
  }\href@noop {} {\bibfield  {journal} {\bibinfo  {journal} {International
  Journal of Smart and Nano Materials}\ }\textbf {\bibinfo {volume} {15}},\
  \bibinfo {pages} {165} (\bibinfo {year} {2024})}\BibitemShut {NoStop}%
\bibitem [{\citenamefont {Trick}\ \emph {et~al.}(2017)\citenamefont {Trick},
  \citenamefont {Song}, \citenamefont {Wallace},\ and\ \citenamefont
  {Sansom}}]{trick2017}%
  \BibitemOpen
  \bibfield  {author} {\bibinfo {author} {\bibfnamefont {J.~L.}\ \bibnamefont
  {Trick}}, \bibinfo {author} {\bibfnamefont {C.}~\bibnamefont {Song}},
  \bibinfo {author} {\bibfnamefont {E.~J.}\ \bibnamefont {Wallace}}, \ and\
  \bibinfo {author} {\bibfnamefont {M.~S.}\ \bibnamefont {Sansom}},\
  }\href@noop {} {\bibfield  {journal} {\bibinfo  {journal} {ACS nano}\
  }\textbf {\bibinfo {volume} {11}},\ \bibinfo {pages} {1840} (\bibinfo {year}
  {2017})}\BibitemShut {NoStop}%
\bibitem [{\citenamefont {Paulo}\ \emph
  {et~al.}(2023{\natexlab{b}})\citenamefont {Paulo}, \citenamefont {Gubbiotti},
  \citenamefont {Grosu}, \citenamefont {Meloni},\ and\ \citenamefont
  {Giacomello}}]{Paulo2023}%
  \BibitemOpen
  \bibfield  {author} {\bibinfo {author} {\bibfnamefont {G.}~\bibnamefont
  {Paulo}}, \bibinfo {author} {\bibfnamefont {A.}~\bibnamefont {Gubbiotti}},
  \bibinfo {author} {\bibfnamefont {Y.}~\bibnamefont {Grosu}}, \bibinfo
  {author} {\bibfnamefont {S.}~\bibnamefont {Meloni}}, \ and\ \bibinfo {author}
  {\bibfnamefont {A.}~\bibnamefont {Giacomello}},\ }\href {\doibase
  10.1038/s42005-023-01140-0} {\bibfield  {journal} {\bibinfo  {journal}
  {Communications Physics}\ }\textbf {\bibinfo {volume} {6}} (\bibinfo {year}
  {2023}{\natexlab{b}}),\ 10.1038/s42005-023-01140-0}\BibitemShut {NoStop}%
\bibitem [{\citenamefont {Bonthuis}\ \emph {et~al.}(2011)\citenamefont
  {Bonthuis}, \citenamefont {Gekle},\ and\ \citenamefont
  {Netz}}]{bonthuis2011dielectric}%
  \BibitemOpen
  \bibfield  {author} {\bibinfo {author} {\bibfnamefont {D.~J.}\ \bibnamefont
  {Bonthuis}}, \bibinfo {author} {\bibfnamefont {S.}~\bibnamefont {Gekle}}, \
  and\ \bibinfo {author} {\bibfnamefont {R.~R.}\ \bibnamefont {Netz}},\
  }\href@noop {} {\bibfield  {journal} {\bibinfo  {journal} {Physical review
  letters}\ }\textbf {\bibinfo {volume} {107}},\ \bibinfo {pages} {166102}
  (\bibinfo {year} {2011})}\BibitemShut {NoStop}%
\bibitem [{\citenamefont {Schlaich}\ \emph {et~al.}(2016)\citenamefont
  {Schlaich}, \citenamefont {Knapp},\ and\ \citenamefont
  {Netz}}]{schlaich2016water}%
  \BibitemOpen
  \bibfield  {author} {\bibinfo {author} {\bibfnamefont {A.}~\bibnamefont
  {Schlaich}}, \bibinfo {author} {\bibfnamefont {E.~W.}\ \bibnamefont {Knapp}},
  \ and\ \bibinfo {author} {\bibfnamefont {R.~R.}\ \bibnamefont {Netz}},\
  }\href@noop {} {\bibfield  {journal} {\bibinfo  {journal} {Physical review
  letters}\ }\textbf {\bibinfo {volume} {117}},\ \bibinfo {pages} {048001}
  (\bibinfo {year} {2016})}\BibitemShut {NoStop}%
\bibitem [{\citenamefont {Becker}\ and\ \citenamefont
  {Netz}(2024)}]{becker2024interfacial}%
  \BibitemOpen
  \bibfield  {author} {\bibinfo {author} {\bibfnamefont {M.~R.}\ \bibnamefont
  {Becker}}\ and\ \bibinfo {author} {\bibfnamefont {R.~R.}\ \bibnamefont
  {Netz}},\ }\href@noop {} {\bibfield  {journal} {\bibinfo  {journal} {The
  Journal of Chemical Physics}\ }\textbf {\bibinfo {volume} {161}} (\bibinfo
  {year} {2024})}\BibitemShut {NoStop}%
\bibitem [{\citenamefont {Kirby}(2010)}]{kirby2010micro}%
  \BibitemOpen
  \bibfield  {author} {\bibinfo {author} {\bibfnamefont {B.~J.}\ \bibnamefont
  {Kirby}},\ }\href@noop {} {\emph {\bibinfo {title} {Micro-and nanoscale fluid
  mechanics: transport in microfluidic devices}}}\ (\bibinfo  {publisher}
  {Cambridge university press},\ \bibinfo {year} {2010})\BibitemShut {NoStop}%
\bibitem [{\citenamefont {Humphrey}\ \emph {et~al.}(1996)\citenamefont
  {Humphrey}, \citenamefont {Dalke},\ and\ \citenamefont {Schulten}}]{vmd}%
  \BibitemOpen
  \bibfield  {author} {\bibinfo {author} {\bibfnamefont {W.}~\bibnamefont
  {Humphrey}}, \bibinfo {author} {\bibfnamefont {A.}~\bibnamefont {Dalke}}, \
  and\ \bibinfo {author} {\bibfnamefont {K.}~\bibnamefont {Schulten}},\
  }\href@noop {} {\bibfield  {journal} {\bibinfo  {journal} {Journal of
  molecular graphics}\ }\textbf {\bibinfo {volume} {14}},\ \bibinfo {pages}
  {33} (\bibinfo {year} {1996})}\BibitemShut {NoStop}%
\bibitem [{\citenamefont {Paulo}\ \emph
  {et~al.}(2023{\natexlab{c}})\citenamefont {Paulo}, \citenamefont
  {Gubbiotti},\ and\ \citenamefont {Giacomello}}]{Paulo2023jcp}%
  \BibitemOpen
  \bibfield  {author} {\bibinfo {author} {\bibfnamefont {G.}~\bibnamefont
  {Paulo}}, \bibinfo {author} {\bibfnamefont {A.}~\bibnamefont {Gubbiotti}}, \
  and\ \bibinfo {author} {\bibfnamefont {A.}~\bibnamefont {Giacomello}},\
  }\href {\doibase 10.1063/5.0147647} {\bibfield  {journal} {\bibinfo
  {journal} {The Journal of Chemical Physics}\ }\textbf {\bibinfo {volume}
  {158}} (\bibinfo {year} {2023}{\natexlab{c}}),\
  10.1063/5.0147647}\BibitemShut {NoStop}%
\bibitem [{\citenamefont {Chua}\ \emph {et~al.}(2012)\citenamefont {Chua},
  \citenamefont {Sbitnev},\ and\ \citenamefont {Kim}}]{CHUA2012}%
  \BibitemOpen
  \bibfield  {author} {\bibinfo {author} {\bibfnamefont {L.}~\bibnamefont
  {Chua}}, \bibinfo {author} {\bibfnamefont {V.}~\bibnamefont {Sbitnev}}, \
  and\ \bibinfo {author} {\bibfnamefont {H.}~\bibnamefont {Kim}},\ }\href
  {\doibase 10.1142/s021812741230011x} {\bibfield  {journal} {\bibinfo
  {journal} {International Journal of Bifurcation and Chaos}\ }\textbf
  {\bibinfo {volume} {22}},\ \bibinfo {pages} {1230011} (\bibinfo {year}
  {2012})}\BibitemShut {NoStop}%
\bibitem [{\citenamefont {Sun}\ \emph {et~al.}(2019)\citenamefont {Sun},
  \citenamefont {Chen}, \citenamefont {Xiao}, \citenamefont {Zhou},
  \citenamefont {Ranjan}, \citenamefont {Hou}, \citenamefont {Zhu},
  \citenamefont {Zhao}, \citenamefont {Redfern},\ and\ \citenamefont
  {Zhou}}]{sun2019}%
  \BibitemOpen
  \bibfield  {author} {\bibinfo {author} {\bibfnamefont {B.}~\bibnamefont
  {Sun}}, \bibinfo {author} {\bibfnamefont {Y.}~\bibnamefont {Chen}}, \bibinfo
  {author} {\bibfnamefont {M.}~\bibnamefont {Xiao}}, \bibinfo {author}
  {\bibfnamefont {G.}~\bibnamefont {Zhou}}, \bibinfo {author} {\bibfnamefont
  {S.}~\bibnamefont {Ranjan}}, \bibinfo {author} {\bibfnamefont
  {W.}~\bibnamefont {Hou}}, \bibinfo {author} {\bibfnamefont {X.}~\bibnamefont
  {Zhu}}, \bibinfo {author} {\bibfnamefont {Y.}~\bibnamefont {Zhao}}, \bibinfo
  {author} {\bibfnamefont {S.~A.}\ \bibnamefont {Redfern}}, \ and\ \bibinfo
  {author} {\bibfnamefont {Y.~N.}\ \bibnamefont {Zhou}},\ }\href@noop {}
  {\bibfield  {journal} {\bibinfo  {journal} {Nano letters}\ }\textbf {\bibinfo
  {volume} {19}},\ \bibinfo {pages} {6461} (\bibinfo {year}
  {2019})}\BibitemShut {NoStop}%
\bibitem [{\citenamefont {Cervera}\ \emph {et~al.}(2006)\citenamefont
  {Cervera}, \citenamefont {Schiedt}, \citenamefont {Neumann}, \citenamefont
  {Maf{\'e}},\ and\ \citenamefont {Ram{\'\i}rez}}]{cervera2006ionic}%
  \BibitemOpen
  \bibfield  {author} {\bibinfo {author} {\bibfnamefont {J.}~\bibnamefont
  {Cervera}}, \bibinfo {author} {\bibfnamefont {B.}~\bibnamefont {Schiedt}},
  \bibinfo {author} {\bibfnamefont {R.}~\bibnamefont {Neumann}}, \bibinfo
  {author} {\bibfnamefont {S.}~\bibnamefont {Maf{\'e}}}, \ and\ \bibinfo
  {author} {\bibfnamefont {P.}~\bibnamefont {Ram{\'\i}rez}},\ }\href@noop {}
  {\bibfield  {journal} {\bibinfo  {journal} {The Journal of chemical physics}\
  }\textbf {\bibinfo {volume} {124}} (\bibinfo {year} {2006})}\BibitemShut
  {NoStop}%
\bibitem [{\citenamefont {Robin}\ \emph {et~al.}(2023)\citenamefont {Robin},
  \citenamefont {Emmerich}, \citenamefont {Ismail}, \citenamefont {Nigu{\`e}s},
  \citenamefont {You}, \citenamefont {Nam}, \citenamefont {Keerthi},
  \citenamefont {Siria}, \citenamefont {Geim}, \citenamefont {Radha} \emph
  {et~al.}}]{robin2023long}%
  \BibitemOpen
  \bibfield  {author} {\bibinfo {author} {\bibfnamefont {P.}~\bibnamefont
  {Robin}}, \bibinfo {author} {\bibfnamefont {T.}~\bibnamefont {Emmerich}},
  \bibinfo {author} {\bibfnamefont {A.}~\bibnamefont {Ismail}}, \bibinfo
  {author} {\bibfnamefont {A.}~\bibnamefont {Nigu{\`e}s}}, \bibinfo {author}
  {\bibfnamefont {Y.}~\bibnamefont {You}}, \bibinfo {author} {\bibfnamefont
  {G.-H.}\ \bibnamefont {Nam}}, \bibinfo {author} {\bibfnamefont
  {A.}~\bibnamefont {Keerthi}}, \bibinfo {author} {\bibfnamefont
  {A.}~\bibnamefont {Siria}}, \bibinfo {author} {\bibfnamefont
  {A.}~\bibnamefont {Geim}}, \bibinfo {author} {\bibfnamefont {B.}~\bibnamefont
  {Radha}},  \emph {et~al.},\ }\href@noop {} {\bibfield  {journal} {\bibinfo
  {journal} {Science}\ }\textbf {\bibinfo {volume} {379}},\ \bibinfo {pages}
  {161} (\bibinfo {year} {2023})}\BibitemShut {NoStop}%
\bibitem [{\citenamefont {Ismail}\ \emph {et~al.}(2025)\citenamefont {Ismail},
  \citenamefont {Nam}, \citenamefont {Lokhandwala}, \citenamefont {Pandey},
  \citenamefont {Saurav}, \citenamefont {You}, \citenamefont {Jyothilal},
  \citenamefont {Goutham}, \citenamefont {Sajja}, \citenamefont {Keerthi} \emph
  {et~al.}}]{ismail2025programmable}%
  \BibitemOpen
  \bibfield  {author} {\bibinfo {author} {\bibfnamefont {A.}~\bibnamefont
  {Ismail}}, \bibinfo {author} {\bibfnamefont {G.-H.}\ \bibnamefont {Nam}},
  \bibinfo {author} {\bibfnamefont {A.}~\bibnamefont {Lokhandwala}}, \bibinfo
  {author} {\bibfnamefont {S.~V.}\ \bibnamefont {Pandey}}, \bibinfo {author}
  {\bibfnamefont {K.~V.}\ \bibnamefont {Saurav}}, \bibinfo {author}
  {\bibfnamefont {Y.}~\bibnamefont {You}}, \bibinfo {author} {\bibfnamefont
  {H.}~\bibnamefont {Jyothilal}}, \bibinfo {author} {\bibfnamefont
  {S.}~\bibnamefont {Goutham}}, \bibinfo {author} {\bibfnamefont
  {R.}~\bibnamefont {Sajja}}, \bibinfo {author} {\bibfnamefont
  {A.}~\bibnamefont {Keerthi}},  \emph {et~al.},\ }\href@noop {} {\bibfield
  {journal} {\bibinfo  {journal} {Nature Communications}\ }\textbf {\bibinfo
  {volume} {16}},\ \bibinfo {pages} {7008} (\bibinfo {year}
  {2025})}\BibitemShut {NoStop}%
\bibitem [{\citenamefont {Hodgkin}\ and\ \citenamefont
  {Huxley}(1952)}]{hodgkin1952quantitative}%
  \BibitemOpen
  \bibfield  {author} {\bibinfo {author} {\bibfnamefont {A.~L.}\ \bibnamefont
  {Hodgkin}}\ and\ \bibinfo {author} {\bibfnamefont {A.~F.}\ \bibnamefont
  {Huxley}},\ }\href@noop {} {\bibfield  {journal} {\bibinfo  {journal} {The
  Journal of physiology}\ }\textbf {\bibinfo {volume} {117}},\ \bibinfo {pages}
  {500} (\bibinfo {year} {1952})}\BibitemShut {NoStop}%
\bibitem [{\citenamefont {Chua}\ \emph {et~al.}(1983)\citenamefont {Chua},
  \citenamefont {Yu},\ and\ \citenamefont {Yu}}]{chua1983negative}%
  \BibitemOpen
  \bibfield  {author} {\bibinfo {author} {\bibfnamefont {L.~O.}\ \bibnamefont
  {Chua}}, \bibinfo {author} {\bibfnamefont {J.}~\bibnamefont {Yu}}, \ and\
  \bibinfo {author} {\bibfnamefont {Y.}~\bibnamefont {Yu}},\ }\href@noop {}
  {\bibfield  {journal} {\bibinfo  {journal} {International Journal of Circuit
  Theory and Applications}\ }\textbf {\bibinfo {volume} {11}},\ \bibinfo
  {pages} {161} (\bibinfo {year} {1983})}\BibitemShut {NoStop}%
\bibitem [{\citenamefont {Berger}\ and\ \citenamefont
  {Ramesh}(2011)}]{berger2011chapter5}%
  \BibitemOpen
  \bibfield  {author} {\bibinfo {author} {\bibfnamefont {P.~R.}\ \bibnamefont
  {Berger}}\ and\ \bibinfo {author} {\bibfnamefont {A.}~\bibnamefont
  {Ramesh}},\ }in\ \href@noop {} {\emph {\bibinfo {booktitle} {Comprehensive
  Semiconductor Science and Technology}}}\ (\bibinfo  {publisher} {Elsevier},\
  \bibinfo {year} {2011})\ Chap.~\bibinfo {chapter} {5}, pp.\ \bibinfo {pages}
  {176--241}\BibitemShut {NoStop}%
\bibitem [{\citenamefont {Sauciuc}\ \emph
  {et~al.}(2024{\natexlab{a}})\citenamefont {Sauciuc}, \citenamefont
  {Morozzo~della Rocca}, \citenamefont {Tadema}, \citenamefont {Chinappi},\
  and\ \citenamefont {Maglia}}]{sauciuc2024translocation}%
  \BibitemOpen
  \bibfield  {author} {\bibinfo {author} {\bibfnamefont {A.}~\bibnamefont
  {Sauciuc}}, \bibinfo {author} {\bibfnamefont {B.}~\bibnamefont {Morozzo~della
  Rocca}}, \bibinfo {author} {\bibfnamefont {M.~J.}\ \bibnamefont {Tadema}},
  \bibinfo {author} {\bibfnamefont {M.}~\bibnamefont {Chinappi}}, \ and\
  \bibinfo {author} {\bibfnamefont {G.}~\bibnamefont {Maglia}},\ }\href@noop {}
  {\bibfield  {journal} {\bibinfo  {journal} {Nature Biotechnology}\ }\textbf
  {\bibinfo {volume} {42}},\ \bibinfo {pages} {1275} (\bibinfo {year}
  {2024}{\natexlab{a}})}\BibitemShut {NoStop}%
\bibitem [{\citenamefont {Sauciuc}\ \emph
  {et~al.}(2024{\natexlab{b}})\citenamefont {Sauciuc}, \citenamefont
  {Whittaker}, \citenamefont {Tadema}, \citenamefont {Tych}, \citenamefont
  {Guskov},\ and\ \citenamefont {Maglia}}]{sauciuc2024blobs}%
  \BibitemOpen
  \bibfield  {author} {\bibinfo {author} {\bibfnamefont {A.}~\bibnamefont
  {Sauciuc}}, \bibinfo {author} {\bibfnamefont {J.}~\bibnamefont {Whittaker}},
  \bibinfo {author} {\bibfnamefont {M.}~\bibnamefont {Tadema}}, \bibinfo
  {author} {\bibfnamefont {K.}~\bibnamefont {Tych}}, \bibinfo {author}
  {\bibfnamefont {A.}~\bibnamefont {Guskov}}, \ and\ \bibinfo {author}
  {\bibfnamefont {G.}~\bibnamefont {Maglia}},\ }\href@noop {} {\bibfield
  {journal} {\bibinfo  {journal} {Proceedings of the National Academy of
  Sciences}\ }\textbf {\bibinfo {volume} {121}},\ \bibinfo {pages}
  {e2405018121} (\bibinfo {year} {2024}{\natexlab{b}})}\BibitemShut {NoStop}%
\bibitem [{\citenamefont {Versloot}\ \emph {et~al.}(2022)\citenamefont
  {Versloot}, \citenamefont {Straathof}, \citenamefont {Stouwie}, \citenamefont
  {Tadema},\ and\ \citenamefont {Maglia}}]{versloot2022beta}%
  \BibitemOpen
  \bibfield  {author} {\bibinfo {author} {\bibfnamefont {R.~C.~A.}\
  \bibnamefont {Versloot}}, \bibinfo {author} {\bibfnamefont {S.~A.~P.}\
  \bibnamefont {Straathof}}, \bibinfo {author} {\bibfnamefont {G.}~\bibnamefont
  {Stouwie}}, \bibinfo {author} {\bibfnamefont {M.~J.}\ \bibnamefont {Tadema}},
  \ and\ \bibinfo {author} {\bibfnamefont {G.}~\bibnamefont {Maglia}},\
  }\href@noop {} {\bibfield  {journal} {\bibinfo  {journal} {ACS nano}\
  }\textbf {\bibinfo {volume} {16}},\ \bibinfo {pages} {7258} (\bibinfo {year}
  {2022})}\BibitemShut {NoStop}%
\bibitem [{\citenamefont {Siwy}(2006)}]{siwy2006ion}%
  \BibitemOpen
  \bibfield  {author} {\bibinfo {author} {\bibfnamefont {Z.~S.}\ \bibnamefont
  {Siwy}},\ }\href@noop {} {\bibfield  {journal} {\bibinfo  {journal} {Advanced
  Functional Materials}\ }\textbf {\bibinfo {volume} {16}},\ \bibinfo {pages}
  {735} (\bibinfo {year} {2006})}\BibitemShut {NoStop}%
\bibitem [{\citenamefont {Ramirez}\ \emph {et~al.}(2008)\citenamefont
  {Ramirez}, \citenamefont {Apel}, \citenamefont {Cervera},\ and\ \citenamefont
  {Maf{\'e}}}]{ramirez2008pore}%
  \BibitemOpen
  \bibfield  {author} {\bibinfo {author} {\bibfnamefont {P.}~\bibnamefont
  {Ramirez}}, \bibinfo {author} {\bibfnamefont {P.~Y.}\ \bibnamefont {Apel}},
  \bibinfo {author} {\bibfnamefont {J.}~\bibnamefont {Cervera}}, \ and\
  \bibinfo {author} {\bibfnamefont {S.}~\bibnamefont {Maf{\'e}}},\ }\href@noop
  {} {\bibfield  {journal} {\bibinfo  {journal} {Nanotechnology}\ }\textbf
  {\bibinfo {volume} {19}},\ \bibinfo {pages} {315707} (\bibinfo {year}
  {2008})}\BibitemShut {NoStop}%
\bibitem [{\citenamefont {Laucirica}\ \emph {et~al.}(2021)\citenamefont
  {Laucirica}, \citenamefont {Toimil-Molares}, \citenamefont {Trautmann},
  \citenamefont {Marmisoll{\'e}},\ and\ \citenamefont
  {Azzaroni}}]{laucirica2021nanofluidic}%
  \BibitemOpen
  \bibfield  {author} {\bibinfo {author} {\bibfnamefont {G.}~\bibnamefont
  {Laucirica}}, \bibinfo {author} {\bibfnamefont {M.~E.}\ \bibnamefont
  {Toimil-Molares}}, \bibinfo {author} {\bibfnamefont {C.}~\bibnamefont
  {Trautmann}}, \bibinfo {author} {\bibfnamefont {W.}~\bibnamefont
  {Marmisoll{\'e}}}, \ and\ \bibinfo {author} {\bibfnamefont {O.}~\bibnamefont
  {Azzaroni}},\ }\href@noop {} {\bibfield  {journal} {\bibinfo  {journal}
  {Chemical Science}\ }\textbf {\bibinfo {volume} {12}},\ \bibinfo {pages}
  {12874} (\bibinfo {year} {2021})}\BibitemShut {NoStop}%
\bibitem [{\citenamefont {Ai}\ \emph {et~al.}(2010)\citenamefont {Ai},
  \citenamefont {Zhang}, \citenamefont {Joo}, \citenamefont {Cheney},\ and\
  \citenamefont {Qian}}]{ai2010effects}%
  \BibitemOpen
  \bibfield  {author} {\bibinfo {author} {\bibfnamefont {Y.}~\bibnamefont
  {Ai}}, \bibinfo {author} {\bibfnamefont {M.}~\bibnamefont {Zhang}}, \bibinfo
  {author} {\bibfnamefont {S.~W.}\ \bibnamefont {Joo}}, \bibinfo {author}
  {\bibfnamefont {M.~A.}\ \bibnamefont {Cheney}}, \ and\ \bibinfo {author}
  {\bibfnamefont {S.}~\bibnamefont {Qian}},\ }\href@noop {} {\bibfield
  {journal} {\bibinfo  {journal} {The Journal of Physical Chemistry C}\
  }\textbf {\bibinfo {volume} {114}},\ \bibinfo {pages} {3883} (\bibinfo {year}
  {2010})}\BibitemShut {NoStop}%
\bibitem [{\citenamefont {Baldelli}\ \emph
  {et~al.}(2024{\natexlab{a}})\citenamefont {Baldelli}, \citenamefont
  {Di~Muccio}, \citenamefont {Viola}, \citenamefont {Giacomello}, \citenamefont
  {Cecconi}, \citenamefont {Balme},\ and\ \citenamefont
  {Chinappi}}]{baldelli2024performance}%
  \BibitemOpen
  \bibfield  {author} {\bibinfo {author} {\bibfnamefont {M.}~\bibnamefont
  {Baldelli}}, \bibinfo {author} {\bibfnamefont {G.}~\bibnamefont {Di~Muccio}},
  \bibinfo {author} {\bibfnamefont {F.}~\bibnamefont {Viola}}, \bibinfo
  {author} {\bibfnamefont {A.}~\bibnamefont {Giacomello}}, \bibinfo {author}
  {\bibfnamefont {F.}~\bibnamefont {Cecconi}}, \bibinfo {author} {\bibfnamefont
  {S.}~\bibnamefont {Balme}}, \ and\ \bibinfo {author} {\bibfnamefont
  {M.}~\bibnamefont {Chinappi}},\ }\href@noop {} {\bibfield  {journal}
  {\bibinfo  {journal} {ChemPhysChem}\ }\textbf {\bibinfo {volume} {25}},\
  \bibinfo {pages} {e202400395} (\bibinfo {year}
  {2024}{\natexlab{a}})}\BibitemShut {NoStop}%
\bibitem [{\citenamefont {Hille}(1970)}]{hille1970ionic}%
  \BibitemOpen
  \bibfield  {author} {\bibinfo {author} {\bibfnamefont {B.}~\bibnamefont
  {Hille}},\ }\href@noop {} {\bibfield  {journal} {\bibinfo  {journal}
  {Progress in biophysics and molecular biology}\ }\textbf {\bibinfo {volume}
  {21}},\ \bibinfo {pages} {1} (\bibinfo {year} {1970})}\BibitemShut {NoStop}%
\bibitem [{\citenamefont {Tourneur}(1986)}]{Tourneur1986}%
  \BibitemOpen
  \bibfield  {author} {\bibinfo {author} {\bibfnamefont {Y.}~\bibnamefont
  {Tourneur}},\ }\href {\doibase doi.org/10.1007/BF01869929} {\bibfield
  {journal} {\bibinfo  {journal} {Journal of Membrane Biology}\ }\textbf
  {\bibinfo {volume} {90}},\ \bibinfo {pages} {115–122} (\bibinfo {year}
  {1986})}\BibitemShut {NoStop}%
\bibitem [{\citenamefont {Berendsen}\ \emph {et~al.}(1987)\citenamefont
  {Berendsen}, \citenamefont {Grigera},\ and\ \citenamefont
  {Straatsma}}]{berendsen1987missing}%
  \BibitemOpen
  \bibfield  {author} {\bibinfo {author} {\bibfnamefont {H.}~\bibnamefont
  {Berendsen}}, \bibinfo {author} {\bibfnamefont {J.}~\bibnamefont {Grigera}},
  \ and\ \bibinfo {author} {\bibfnamefont {T.}~\bibnamefont {Straatsma}},\
  }\href@noop {} {\bibfield  {journal} {\bibinfo  {journal} {Journal of
  Physical Chemistry}\ }\textbf {\bibinfo {volume} {91}},\ \bibinfo {pages}
  {6269} (\bibinfo {year} {1987})}\BibitemShut {NoStop}%
\bibitem [{\citenamefont {Camisasca}\ \emph {et~al.}(2020)\citenamefont
  {Camisasca}, \citenamefont {Tinti},\ and\ \citenamefont
  {Giacomello}}]{camisasca2020gas}%
  \BibitemOpen
  \bibfield  {author} {\bibinfo {author} {\bibfnamefont {G.}~\bibnamefont
  {Camisasca}}, \bibinfo {author} {\bibfnamefont {A.}~\bibnamefont {Tinti}}, \
  and\ \bibinfo {author} {\bibfnamefont {A.}~\bibnamefont {Giacomello}},\
  }\href@noop {} {\bibfield  {journal} {\bibinfo  {journal} {The Journal of
  Physical Chemistry Letters}\ }\textbf {\bibinfo {volume} {11}},\ \bibinfo
  {pages} {9171} (\bibinfo {year} {2020})}\BibitemShut {NoStop}%
\bibitem [{\citenamefont {Marchio}\ \emph {et~al.}(2018)\citenamefont
  {Marchio}, \citenamefont {Meloni}, \citenamefont {Giacomello}, \citenamefont
  {Valeriani},\ and\ \citenamefont {Casciola}}]{marchio2018}%
  \BibitemOpen
  \bibfield  {author} {\bibinfo {author} {\bibfnamefont {S.}~\bibnamefont
  {Marchio}}, \bibinfo {author} {\bibfnamefont {S.}~\bibnamefont {Meloni}},
  \bibinfo {author} {\bibfnamefont {A.}~\bibnamefont {Giacomello}}, \bibinfo
  {author} {\bibfnamefont {C.}~\bibnamefont {Valeriani}}, \ and\ \bibinfo
  {author} {\bibfnamefont {C.}~\bibnamefont {Casciola}},\ }\href@noop {}
  {\bibfield  {journal} {\bibinfo  {journal} {The Journal of chemical physics}\
  }\textbf {\bibinfo {volume} {148}},\ \bibinfo {pages} {064706} (\bibinfo
  {year} {2018})}\BibitemShut {NoStop}%
\bibitem [{\citenamefont {Martyna}\ \emph {et~al.}(1992)\citenamefont
  {Martyna}, \citenamefont {Klein},\ and\ \citenamefont
  {Tuckerman}}]{martyna1992}%
  \BibitemOpen
  \bibfield  {author} {\bibinfo {author} {\bibfnamefont {G.~J.}\ \bibnamefont
  {Martyna}}, \bibinfo {author} {\bibfnamefont {M.~L.}\ \bibnamefont {Klein}},
  \ and\ \bibinfo {author} {\bibfnamefont {M.}~\bibnamefont {Tuckerman}},\
  }\href@noop {} {\bibfield  {journal} {\bibinfo  {journal} {The Journal of
  chemical physics}\ }\textbf {\bibinfo {volume} {97}},\ \bibinfo {pages}
  {2635} (\bibinfo {year} {1992})}\BibitemShut {NoStop}%
\bibitem [{\citenamefont {Gumbart}\ \emph {et~al.}(2012)\citenamefont
  {Gumbart}, \citenamefont {Khalili-Araghi}, \citenamefont {Sotomayor},\ and\
  \citenamefont {Roux}}]{gumbart2012constant}%
  \BibitemOpen
  \bibfield  {author} {\bibinfo {author} {\bibfnamefont {J.}~\bibnamefont
  {Gumbart}}, \bibinfo {author} {\bibfnamefont {F.}~\bibnamefont
  {Khalili-Araghi}}, \bibinfo {author} {\bibfnamefont {M.}~\bibnamefont
  {Sotomayor}}, \ and\ \bibinfo {author} {\bibfnamefont {B.}~\bibnamefont
  {Roux}},\ }\href@noop {} {\bibfield  {journal} {\bibinfo  {journal}
  {Biochimica et Biophysica Acta (BBA)-Biomembranes}\ }\textbf {\bibinfo
  {volume} {1818}},\ \bibinfo {pages} {294} (\bibinfo {year}
  {2012})}\BibitemShut {NoStop}%
\bibitem [{\citenamefont {Sauciuc}\ and\ \citenamefont
  {Maglia}(2024)}]{sauciuc2024controlled}%
  \BibitemOpen
  \bibfield  {author} {\bibinfo {author} {\bibfnamefont {A.}~\bibnamefont
  {Sauciuc}}\ and\ \bibinfo {author} {\bibfnamefont {G.}~\bibnamefont
  {Maglia}},\ }\href@noop {} {\bibfield  {journal} {\bibinfo  {journal} {Nano
  Letters}\ }\textbf {\bibinfo {volume} {24}},\ \bibinfo {pages} {14118}
  (\bibinfo {year} {2024})}\BibitemShut {NoStop}%
\bibitem [{\citenamefont {Olsson}\ \emph {et~al.}(2011)\citenamefont {Olsson},
  \citenamefont {S{\o}ndergaard}, \citenamefont {Rostkowski},\ and\
  \citenamefont {Jensen}}]{olsson2011propka3}%
  \BibitemOpen
  \bibfield  {author} {\bibinfo {author} {\bibfnamefont {M.~H.}\ \bibnamefont
  {Olsson}}, \bibinfo {author} {\bibfnamefont {C.~R.}\ \bibnamefont
  {S{\o}ndergaard}}, \bibinfo {author} {\bibfnamefont {M.}~\bibnamefont
  {Rostkowski}}, \ and\ \bibinfo {author} {\bibfnamefont {J.~H.}\ \bibnamefont
  {Jensen}},\ }\href@noop {} {\bibfield  {journal} {\bibinfo  {journal}
  {Journal of chemical theory and computation}\ }\textbf {\bibinfo {volume}
  {7}},\ \bibinfo {pages} {525} (\bibinfo {year} {2011})}\BibitemShut {NoStop}%
\bibitem [{\citenamefont {Debiec}\ \emph {et~al.}(2016)\citenamefont {Debiec},
  \citenamefont {Cerutti}, \citenamefont {Baker}, \citenamefont {Gronenborn},
  \citenamefont {Case},\ and\ \citenamefont {Chong}}]{debiec2016further}%
  \BibitemOpen
  \bibfield  {author} {\bibinfo {author} {\bibfnamefont {K.~T.}\ \bibnamefont
  {Debiec}}, \bibinfo {author} {\bibfnamefont {D.~S.}\ \bibnamefont {Cerutti}},
  \bibinfo {author} {\bibfnamefont {L.~R.}\ \bibnamefont {Baker}}, \bibinfo
  {author} {\bibfnamefont {A.~M.}\ \bibnamefont {Gronenborn}}, \bibinfo
  {author} {\bibfnamefont {D.~A.}\ \bibnamefont {Case}}, \ and\ \bibinfo
  {author} {\bibfnamefont {L.~T.}\ \bibnamefont {Chong}},\ }\href@noop {}
  {\bibfield  {journal} {\bibinfo  {journal} {Journal of chemical theory and
  computation}\ }\textbf {\bibinfo {volume} {12}},\ \bibinfo {pages} {3926}
  (\bibinfo {year} {2016})}\BibitemShut {NoStop}%
\bibitem [{\citenamefont {Dickson}\ \emph {et~al.}(2014)\citenamefont
  {Dickson}, \citenamefont {Madej}, \citenamefont {Skjevik}, \citenamefont
  {Betz}, \citenamefont {Teigen}, \citenamefont {Gould},\ and\ \citenamefont
  {Walker}}]{dickson2014lipid14}%
  \BibitemOpen
  \bibfield  {author} {\bibinfo {author} {\bibfnamefont {C.~J.}\ \bibnamefont
  {Dickson}}, \bibinfo {author} {\bibfnamefont {B.~D.}\ \bibnamefont {Madej}},
  \bibinfo {author} {\bibfnamefont {{\AA}.~A.}\ \bibnamefont {Skjevik}},
  \bibinfo {author} {\bibfnamefont {R.~M.}\ \bibnamefont {Betz}}, \bibinfo
  {author} {\bibfnamefont {K.}~\bibnamefont {Teigen}}, \bibinfo {author}
  {\bibfnamefont {I.~R.}\ \bibnamefont {Gould}}, \ and\ \bibinfo {author}
  {\bibfnamefont {R.~C.}\ \bibnamefont {Walker}},\ }\href@noop {} {\bibfield
  {journal} {\bibinfo  {journal} {Journal of chemical theory and computation}\
  }\textbf {\bibinfo {volume} {10}},\ \bibinfo {pages} {865} (\bibinfo {year}
  {2014})}\BibitemShut {NoStop}%
\bibitem [{\citenamefont {Takemura}\ and\ \citenamefont
  {Kitao}(2012)}]{takemura2012water}%
  \BibitemOpen
  \bibfield  {author} {\bibinfo {author} {\bibfnamefont {K.}~\bibnamefont
  {Takemura}}\ and\ \bibinfo {author} {\bibfnamefont {A.}~\bibnamefont
  {Kitao}},\ }\href@noop {} {\bibfield  {journal} {\bibinfo  {journal} {The
  Journal of Physical Chemistry B}\ }\textbf {\bibinfo {volume} {116}},\
  \bibinfo {pages} {6279} (\bibinfo {year} {2012})}\BibitemShut {NoStop}%
\bibitem [{\citenamefont {Baldelli}\ \emph
  {et~al.}(2024{\natexlab{b}})\citenamefont {Baldelli}, \citenamefont
  {Di~Muccio}, \citenamefont {Sauciuc}, \citenamefont {Morozzo~della Rocca},
  \citenamefont {Viola}, \citenamefont {Balme}, \citenamefont {Bonini},
  \citenamefont {Maglia},\ and\ \citenamefont
  {Chinappi}}]{baldelli2024controlling}%
  \BibitemOpen
  \bibfield  {author} {\bibinfo {author} {\bibfnamefont {M.}~\bibnamefont
  {Baldelli}}, \bibinfo {author} {\bibfnamefont {G.}~\bibnamefont {Di~Muccio}},
  \bibinfo {author} {\bibfnamefont {A.}~\bibnamefont {Sauciuc}}, \bibinfo
  {author} {\bibfnamefont {B.}~\bibnamefont {Morozzo~della Rocca}}, \bibinfo
  {author} {\bibfnamefont {F.}~\bibnamefont {Viola}}, \bibinfo {author}
  {\bibfnamefont {S.}~\bibnamefont {Balme}}, \bibinfo {author} {\bibfnamefont
  {A.}~\bibnamefont {Bonini}}, \bibinfo {author} {\bibfnamefont
  {G.}~\bibnamefont {Maglia}}, \ and\ \bibinfo {author} {\bibfnamefont
  {M.}~\bibnamefont {Chinappi}},\ }\href@noop {} {\bibfield  {journal}
  {\bibinfo  {journal} {Advanced Materials}\ }\textbf {\bibinfo {volume}
  {36}},\ \bibinfo {pages} {2401761} (\bibinfo {year}
  {2024}{\natexlab{b}})}\BibitemShut {NoStop}%
\bibitem [{\citenamefont {Maragliano}\ and\ \citenamefont
  {Vanden-Eijnden}(2006)}]{maragliano2006temperature}%
  \BibitemOpen
  \bibfield  {author} {\bibinfo {author} {\bibfnamefont {L.}~\bibnamefont
  {Maragliano}}\ and\ \bibinfo {author} {\bibfnamefont {E.}~\bibnamefont
  {Vanden-Eijnden}},\ }\href@noop {} {\bibfield  {journal} {\bibinfo  {journal}
  {Chemical physics letters}\ }\textbf {\bibinfo {volume} {426}},\ \bibinfo
  {pages} {168} (\bibinfo {year} {2006})}\BibitemShut {NoStop}%
\bibitem [{\citenamefont {Phillips}\ \emph {et~al.}(2005)\citenamefont
  {Phillips}, \citenamefont {Braun}, \citenamefont {Wang}, \citenamefont
  {Gumbart}, \citenamefont {Tajkhorshid}, \citenamefont {Villa}, \citenamefont
  {Chipot}, \citenamefont {Skeel}, \citenamefont {Kale},\ and\ \citenamefont
  {Schulten}}]{phillips2005scalable}%
  \BibitemOpen
  \bibfield  {author} {\bibinfo {author} {\bibfnamefont {J.~C.}\ \bibnamefont
  {Phillips}}, \bibinfo {author} {\bibfnamefont {R.}~\bibnamefont {Braun}},
  \bibinfo {author} {\bibfnamefont {W.}~\bibnamefont {Wang}}, \bibinfo {author}
  {\bibfnamefont {J.}~\bibnamefont {Gumbart}}, \bibinfo {author} {\bibfnamefont
  {E.}~\bibnamefont {Tajkhorshid}}, \bibinfo {author} {\bibfnamefont
  {E.}~\bibnamefont {Villa}}, \bibinfo {author} {\bibfnamefont
  {C.}~\bibnamefont {Chipot}}, \bibinfo {author} {\bibfnamefont {R.~D.}\
  \bibnamefont {Skeel}}, \bibinfo {author} {\bibfnamefont {L.}~\bibnamefont
  {Kale}}, \ and\ \bibinfo {author} {\bibfnamefont {K.}~\bibnamefont
  {Schulten}},\ }\href@noop {} {\bibfield  {journal} {\bibinfo  {journal}
  {Journal of computational chemistry}\ }\textbf {\bibinfo {volume} {26}},\
  \bibinfo {pages} {1781} (\bibinfo {year} {2005})}\BibitemShut {NoStop}%
\bibitem [{\citenamefont {Fiorin}\ \emph {et~al.}(2013)\citenamefont {Fiorin},
  \citenamefont {Klein},\ and\ \citenamefont {H{\'e}nin}}]{fiorin2013using}%
  \BibitemOpen
  \bibfield  {author} {\bibinfo {author} {\bibfnamefont {G.}~\bibnamefont
  {Fiorin}}, \bibinfo {author} {\bibfnamefont {M.~L.}\ \bibnamefont {Klein}}, \
  and\ \bibinfo {author} {\bibfnamefont {J.}~\bibnamefont {H{\'e}nin}},\
  }\href@noop {} {\bibfield  {journal} {\bibinfo  {journal} {Molecular
  Physics}\ }\textbf {\bibinfo {volume} {111}},\ \bibinfo {pages} {3345}
  (\bibinfo {year} {2013})}\BibitemShut {NoStop}%
\bibitem [{\citenamefont {Fiorin}\ \emph {et~al.}(2020)\citenamefont {Fiorin},
  \citenamefont {Marinelli},\ and\ \citenamefont
  {Faraldo-G{\'o}mez}}]{fiorin2020direct}%
  \BibitemOpen
  \bibfield  {author} {\bibinfo {author} {\bibfnamefont {G.}~\bibnamefont
  {Fiorin}}, \bibinfo {author} {\bibfnamefont {F.}~\bibnamefont {Marinelli}}, \
  and\ \bibinfo {author} {\bibfnamefont {J.~D.}\ \bibnamefont
  {Faraldo-G{\'o}mez}},\ }\href@noop {} {\bibfield  {journal} {\bibinfo
  {journal} {Journal of computational chemistry}\ }\textbf {\bibinfo {volume}
  {41}},\ \bibinfo {pages} {449} (\bibinfo {year} {2020})}\BibitemShut
  {NoStop}%
\bibitem [{\citenamefont {Coronel}\ \emph {et~al.}(2024)\citenamefont
  {Coronel}, \citenamefont {Di~Muccio}, \citenamefont {Rothberg}, \citenamefont
  {Giacomello},\ and\ \citenamefont {Carnevale}}]{coronel2024lipid}%
  \BibitemOpen
  \bibfield  {author} {\bibinfo {author} {\bibfnamefont {L.}~\bibnamefont
  {Coronel}}, \bibinfo {author} {\bibfnamefont {G.}~\bibnamefont {Di~Muccio}},
  \bibinfo {author} {\bibfnamefont {B.}~\bibnamefont {Rothberg}}, \bibinfo
  {author} {\bibfnamefont {A.}~\bibnamefont {Giacomello}}, \ and\ \bibinfo
  {author} {\bibfnamefont {V.}~\bibnamefont {Carnevale}},\ }\href@noop {}
  {\bibfield  {journal} {\bibinfo  {journal} {arXiv preprint arXiv:2405.04644}\
  } (\bibinfo {year} {2024})}\BibitemShut {NoStop}%
\bibitem [{\citenamefont {Maglia}\ \emph {et~al.}(2010)\citenamefont {Maglia},
  \citenamefont {Heron}, \citenamefont {Stoddart}, \citenamefont {Japrung},\
  and\ \citenamefont {Bayley}}]{maglia2010analysis}%
  \BibitemOpen
  \bibfield  {author} {\bibinfo {author} {\bibfnamefont {G.}~\bibnamefont
  {Maglia}}, \bibinfo {author} {\bibfnamefont {A.~J.}\ \bibnamefont {Heron}},
  \bibinfo {author} {\bibfnamefont {D.}~\bibnamefont {Stoddart}}, \bibinfo
  {author} {\bibfnamefont {D.}~\bibnamefont {Japrung}}, \ and\ \bibinfo
  {author} {\bibfnamefont {H.}~\bibnamefont {Bayley}},\ }in\ \href@noop {}
  {\emph {\bibinfo {booktitle} {Methods in enzymology}}},\ Vol.\ \bibinfo
  {volume} {475}\ (\bibinfo  {publisher} {Elsevier},\ \bibinfo {year} {2010})\
  pp.\ \bibinfo {pages} {591--623}\BibitemShut {NoStop}%
\end{thebibliography}%

\clearpage

\newpage
\onecolumngrid
\setcounter{equation}{0}
\setcounter{figure}{0}
\setcounter{table}{0}
\setcounter{page}{1}

\renewcommand{\theequation}{S\arabic{equation}}
\renewcommand{\thefigure}{S\arabic{figure}}
\renewcommand{\thetable}{S\arabic{table}}
\renewcommand{\thepage}{S-\arabic{page}}



\begin{center}
\textbf{\large Supplementary Materials for}\\
\vspace{0.3cm}
\textbf{\large Electrodrying in nanopores: from fundamentals to iontronic and memristive applications}
\\
\vspace{0.5cm}
{Giovanni Di Muccio$^{1,2,\dag}$}
{Gon\c{c}alo Paulo$^{1,\dag}$}
{Lorenzo Iannetti$^{1,\dag}$}
{Adina Sauciuc$^{3,\dag}$}
{Giovanni Maglia$^3$}
{Alberto Giacomello$^{1,\ast}$}
\\
\vspace{0.25cm}
{$^1$ Dipartimento di Ingegneria Meccanica e Aerospaziale, Sapienza Universit\`a di Roma, Rome, Italy}\\
{$^2$ NY-MaSBiC, Dipartimento di Scienze della Vita e dell'Ambiente, Universit\`a Politecnica delle Marche, Ancona, Italy}\\
{$^3$ Groningen Biomolecular Sciences \& Biotechnology Institute,  University of Groningen, Groningen, The Netherlands}
\end{center}

\vspace{1.6 cm}

%
%
%
%
%

\clearpage
\newpage
\section*{Supplementary Note 1. \\
Orientational Free Energy Estimation for a Single Water Molecule}

This note provides an analysis of the orientational free energy change experienced by a single water molecule in an electric field. Water molecules have a permanent dipole moment due to the asymmetric charge distribution between oxygen and hydrogen atoms. When exposed to an electric field, these dipoles tend to align, lowering the electrostatic potential energy energy. The extent of this reduction depends on the electric field's strength relative to the thermal energy.

In nanopores, this interaction critically influences the stability of the water-filled (wet) state. We calculate an ideal orientational free energy change for a single water molecule, deriving the partition function for the weak electric fields case.

The interaction between a dipole moment \( p_w \) and a surrounding electric field \( E_z \) is described by the orientation energy
\[
U = - p_w E_z \cos(\theta) \; ,
\]
where \( \theta \) is the angle between the dipole moment and the electric field. The partition function \( Z \), which accounts for all possible dipole orientations in the field, is given by the integral
\[
Z = \int_0^{2\pi} d\phi \, \int_0^\pi \exp\left(\frac{p_w E_z \cos(\theta)}{k_B T}\right) \sin(\theta) \, d\theta \, 
= 2\pi \int_0^\pi \exp\left(\frac{p_w E_z \cos(\theta)}{k_B T}\right) \sin(\theta) \, d\theta \; 
,
\]
since the interaction energy does not depend on the azimuthal angle \( \phi \), the integral over \( \phi \) gives a factor of \( 2\pi \). 
Letting \( a = \frac{p_w E_z}{k_B T} \), the partition function simplifies to
\[
Z = 2\pi \int_0^\pi \exp(a \cos(\theta)) \sin(\theta) \, d\theta \; 
= 4\pi \cdot \frac{\sinh(a)}{a} \; .
\]
For small \( a \), corresponding to weak electric fields where \( p_w E_z \ll k_B T \), we can expand
\[
\frac{\sinh(a)}{a} \approx 1 + \frac{a^2}{6} + O(a^4)\; .
\]
Substituting into the partition function:
\[
Z \approx 4\pi \left( 1 + \frac{a^2}{6} \right) .
\]
Thus, the Helmholtz free energy is
\[
F = -k_B T \ln Z = -k_B T \ln\left(1 + \left(\frac{p_w E_z}{k_BT}\right)^2 \right) + Const.
\]
with the entropic contribution being
\[
S = -\frac{\partial F}{\partial T} = 
k_B \; \left( 
\ln  \left( 4\pi \left( 1 + \left( \frac{p_w E_z}{k_BT} \right)^2 \right) \right)
- \frac{2 p_w^2 E_z^2}{k_B^2 T^2 + p_w^2 E_z^2} 
\right) \; ,
\]
that, for \( p_w E_z \ll k_B T \), is approximately
\[
S = -2 k_B \; \left( \frac{p_w E_z}{k_B T} \right)^2 + const
\]

Considering \( p_w = 0.6 \, \text{e\AA} \), room temperature \( T = 300 \, \text{K} \), and electric field \( E_z = \frac{V_{int}}{L_z} \), with voltage \( V_{int} = 0.8 \, \text{V} \) across a nanopore of length \( L_z = 3 \, \text{nm} \), we calculate
\[
a = \frac{p_w E_z}{k_B T} = 0.618 \; .
\]
Under these conditions, the total orientational free energy change is
\[
\Delta F = -2.6 k_B T \; .
\]
This reduced free energy favors dipole alignment but is offset by the thermodynamic cost in entropy,
\[
T\Delta S = -0.76 k_B T 
\]
This negative entropy reflects the ordering effect of the electric field as water molecules lose orientational disorder and align.

In conclusion, 
this shows that the presence of an electric field, that can aligns the water dipoles, effectively reduces the system's free energy, stabilizing the wet state.
Indeed, while the alignment of the dipoles decreases the system's entropy by creating a more ordered state, the significant reduction in internal energy could lead to a net decrease in the Helmholtz free energy.

\section*{Supplementary note 2. \\ RLC oscillator model with Negative differential resistance}
The NDR oscillator is realized by modeling an RLC equivalent circuit in parallel with the eletrodrying nanopore of Sec.\ref{sec:MD}. The circuit layout can be seen in the inset of Fig.\ref{fig:applications}b. It's composed of a capacitance C, an inductance L, a resistor R and the electrodrying nanopore. The nanopore is modeled as a memristor $I = g_n \ V$ with a time varying conductance $g_n(t,V)$ \\

$g_n(t,V) = g_0 \ n(t,V)$ \\

with $g_0$ open pore conductance and $n(t,V)$ open pore probability computed from the master equation \\

$\frac{dn}{dt} = (1-n) \ k_w(V) - n \ k_d(V)$ \\

where $k_w(V), \ k_d(V)$ are the wetting and drying rates, respectively, computed as in ref.\cite{paulo2023hydrophobically}. Then the circuit equations for currents and voltage are \\

$I_{ext} = g_0 \ n(V-V_n,t) \left(V-V_n\right) + \frac{1}{R}V \ + C \frac{dV}{dt} \ + \ \frac{1}{L}\int_0^tv(\tau)d\tau$ \\

$\frac{d^2V}{dt^2} = -\frac{1}{C}\left[g_0\left(\frac{dn}{dt}\left(V-V_n\right) \ + \ n\frac{dV}{dt}\right) \ + \ \frac{1}{R}\frac{dV}{dt} \ + \ \frac{1}{L}V \right]$

\clearpage
\newpage
\section*{Supplementary Figure S1. 
\\ The effect of charge on the electrodrying effect }

\begin{figure*}[hb]
    \centering
    \includegraphics[width=1\linewidth]{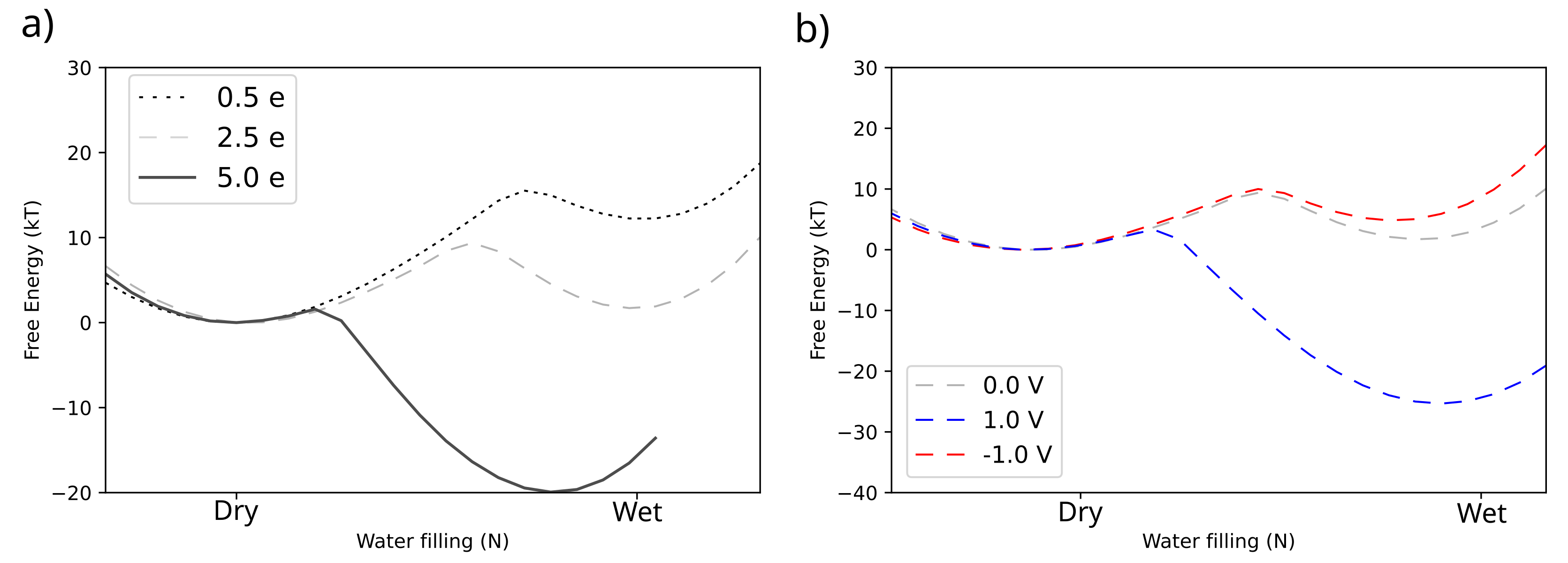}
\caption{Free energy profiles for systems with varying charges at pore entrances. \textbf{(a)} As the magnitude of charge at the pore mouths decreases, the wet state becomes progressively less favorable, diminishing also the electrodrying effect. \textbf{(b)} At a charge magnitude of $2.5e$, the wet and dry states are nearly equally favored, making a full transition between wet and dry states observable under opposite applied voltages, due to the electrodrying asymmetry.
\\\\
In simulations shown in Fig.~2 of the main manuscript, charged pore entrances of $+5e$ and $-5e$ were selected, resulting in a wet pore state at zero voltage that transitions to a dry state upon application of a negative voltage. Here, we examine how variations in pore entrance charges influence this behavior. A decrease in charge magnitude reduces the dipole across the pore, thereby weakening the electrowetting effect. This occurs primarily because the dry state becomes more favorable relative to the wet state at lower charges.
}
    \label{fig:s1}
\end{figure*}

\clearpage
\newpage
\section*{Supplementary Figure S2. \\ The effect of the dipole on the electrodying effect}

\begin{figure}[hb]
    \centering
    \includegraphics[width=0.85\linewidth]{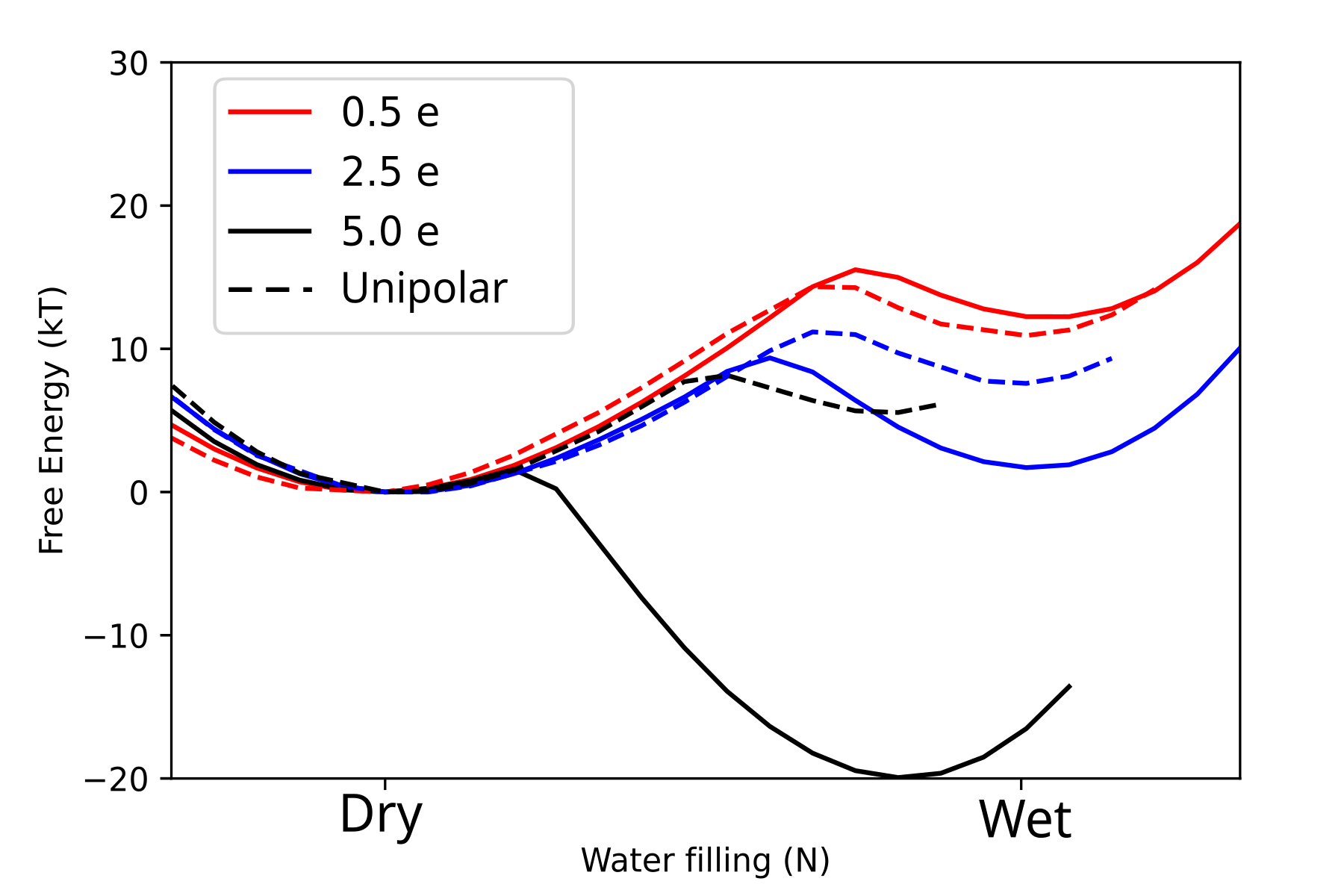}
    \caption{Free energy profiles for dipolar (solid lines) and unipolar (dashed lines) pores at varying charge intensities per ring. Dipolar pores feature oppositely charged rings at the top and bottom entrances, while unipolar pores have identically charged rings. Increasing charge intensity enhances the wetting state stability in dipolar pores, making them suitable candidates for electrodrying. In contrast, unipolar pores remain challenging to wet; their wet state slightly stabilizes but never becomes favorable within the examined charge range. The observed reduction in hydration free energy difference in unipolar pores is likely due to an effective shortening of the hydrophobic region, as charged pore entrances partially increase local hydrophilicity proportional to charge intensity.
    \\\\
    This data further confirms that electrodrying is specifically attributable to the dipole created across the hydrophobic region. Moreover, the intrinsic electric field between oppositely charged pore entrances ensures the wet state at zero voltage, a condition absent in unipolar pores with charges of the same sign.
    }
    \label{fig:s2}
\end{figure}

\clearpage
\newpage
\section*{Supplementary Figure S3. \\ The effect of ions on the electrodying effect}

\begin{figure*}[ht]
    \centering
    \includegraphics[width=0.9\textwidth]{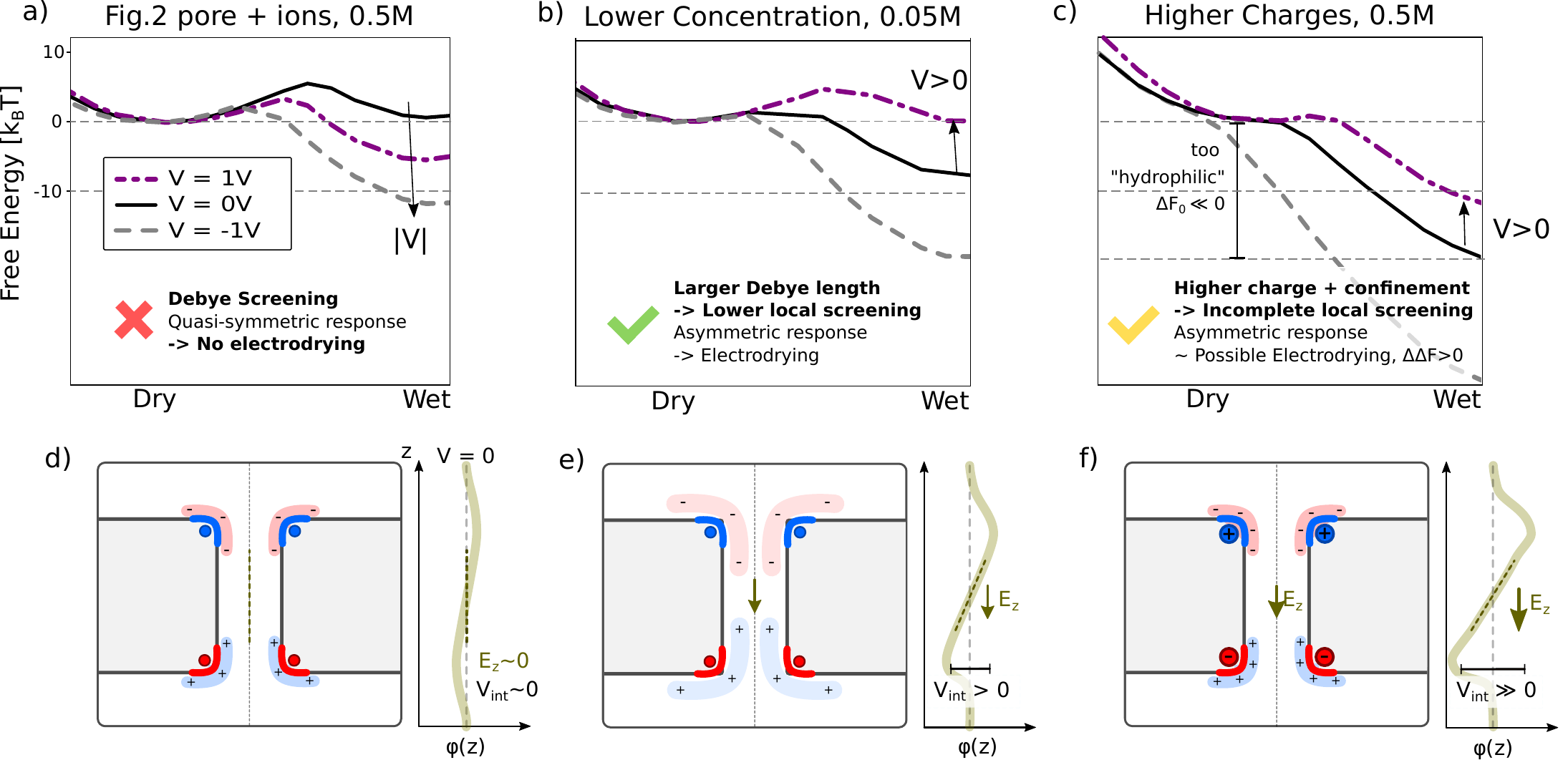}
    \caption{
    \textbf{Effect of ions in electrodrying.} 
    \textbf{a-c)} Free energy profiles for the geometry shown in Fig.\ref{fig:MD-result} immersed into an NaCl electrolyte solution, computed at different charges and salt concentrations.
    \textbf{d-f)} Schematic showing the Debye layers screening the charges at the pore entrance and the resulting electric potential profile, at equilibrium (external $V=0$). When 0.5M NaCl ions are added to the same system considered in Fig.~\ref{fig:MD-result},
    the electrodrying effect disappears, panel a. Indeed, due to the screening counterions clouds, the intrinsic dipole $V_{int}$ across the pore goes to zero. In this scenario, both negative and positive applied voltages wet the pore, resembling the standard electrowetting behaviour.
    Nevertheless, lowering the salt concentration to 0.05M, panel b, restores the electrodrying behaviour, since the Debye length ($\lambda_D = 1.4$nm) is larger than the one needed to completely screen out the intrinsic dipole of the pore. Hence, as for the case without ions, a net $V_{int}$ and an intrinsic electric field $E_z$ at $V=0$ exists, making the pore response to the external electric field asymmetric. Another way to restore the electrodrying behaviour would be to increase the charges at the pore entrance. Also in this case, due to the larger number of counter-ions needed to compensate the pore charges, the screening is not complete and an intrinsic dipole still exist. However, despite the asymmetric $\Delta \Delta F$ electrodrying response is resembled, the equilibrium free energy difference ($\Delta F_0$, see the black line) sees to be too large to be completely reverted by the applied voltage, in the explored voltage range.
    \label{fig:ions}
    }
\end{figure*}

\clearpage
\newpage
\section*{Supplementary Figure S4. 
\\ Illustration of memristor types}
\begin{figure}[hb]
    \centering
    \includegraphics[width=1.0\linewidth]{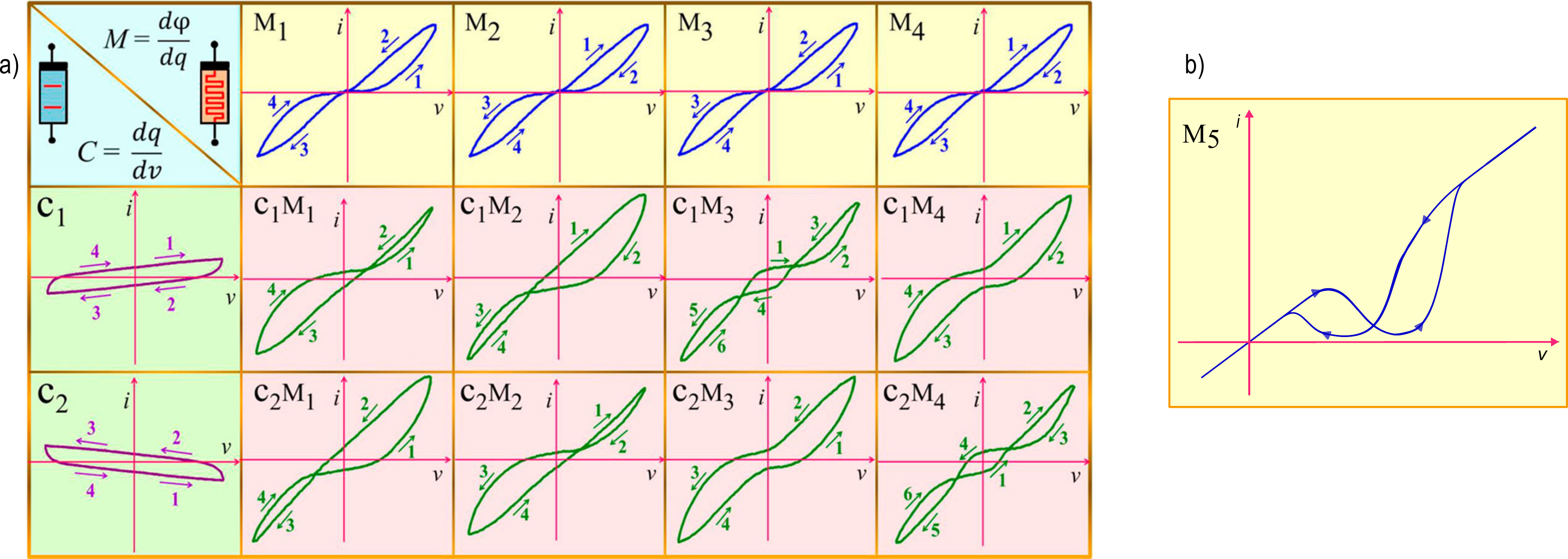}
\caption{Diagram a) from ref.\cite{sun2019}, displaying different I-V hysteresis curves that can be obtained combining four basic types of zero-crossing pinched memristors (M1-M4) with capacitor (C1-C2). Based on the direction of I–V response and the switching direction, this model exhibits eight types of characteristic behaviour not pinched at zero. 
Panel b) shows our electrodrying memristor that, without any capacitance, displays a zero-crossing I-V curve (as M1-M4) with hysteresis not pinched at zero (as for the capacitive-coupled curves, $C_iM_i$). Adapted with permission from "B. Sun, Y. Chen, M. Xiao, G. Zhou, S. Ranjan, W. Hou, X. Zhu,
Y. Zhao, S. A. Redfern, and Y. N. Zhou, Nano letters 19, 6461
(2019)". Copyright 2025 American Chemical Society.
}
\label{fig:memristor_types}
\end{figure}

\clearpage
\newpage
\section*{Supplementary Figure S5. 
\\ Additional CytK mutants conductances for different voltages}
\begin{figure}[hb]
    \centering
    \includegraphics[width=0.85\linewidth]{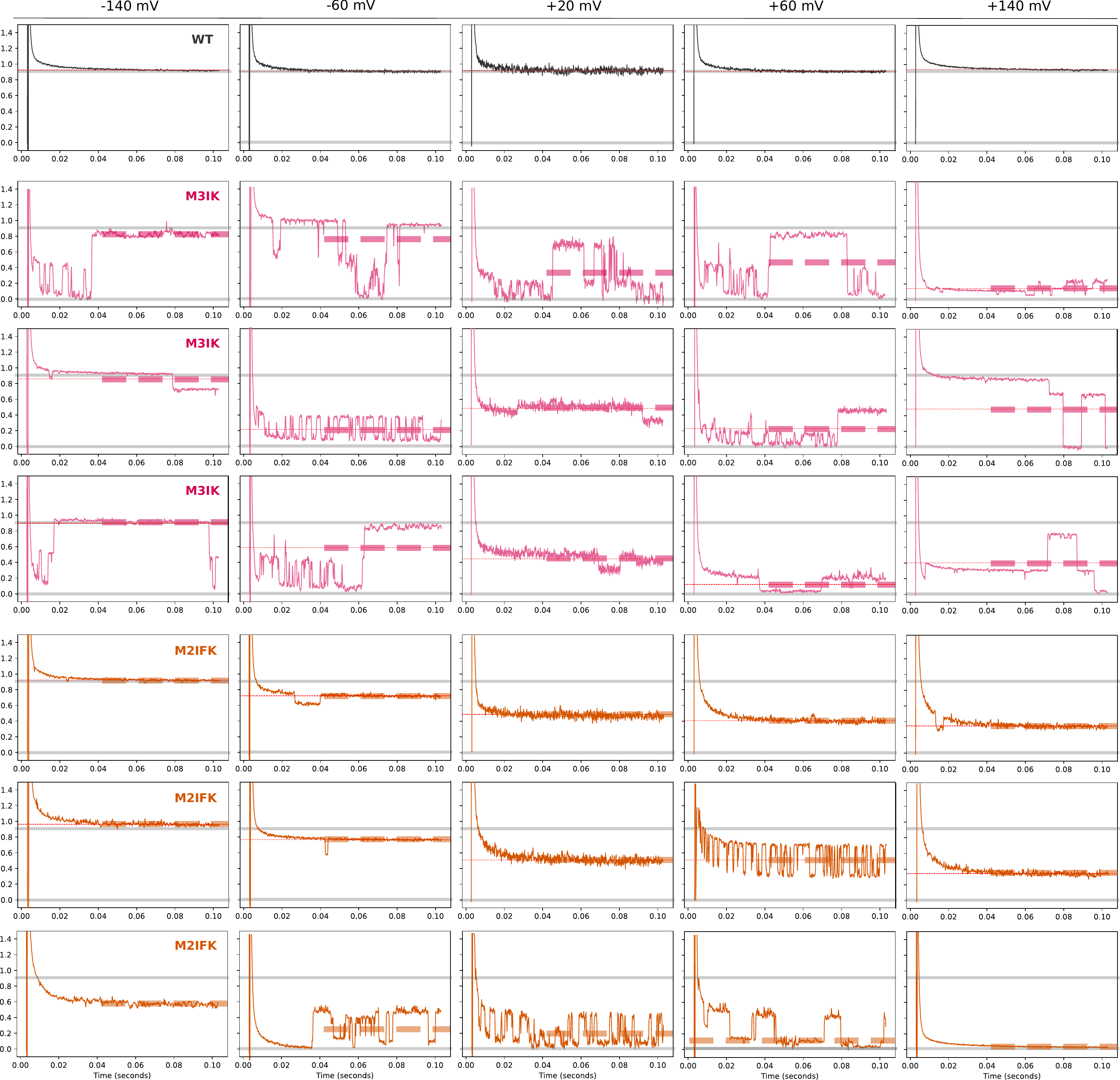}
\caption{\textbf{Conductance profiles of Wild Type CYTK and hydrophobic mutants}. Representative conductance traces (in nS) for WT (top, black) and for mutants M3IK (magenta) and M2IFK (orange) recorded at voltage steps of $-140$~mV, $-60$~mV, $+20$~mV, $+60$~mV, and $+140$~mV. Each row presents traces from a single experiment. The recordings illustrate distinct gating behaviors and variability in conductance stability between the two mutants under the different voltage conditions. Dashed lines in each panel indicate the average conductance computed after 40~ms for the corresponding trace. Grey lines represent zero conductance and the average WT conductance. 
As estimated in previous works~\cite{paulo2023hydrophobically,paulo2024voltage}, the kinetics of wetting and drying can be faster than the recording frequency, with the result that many traces display only the time-averaged conductance between the open and closed states, in accordance with their relative probabilities, and reach zero only when the pore remains fully dry for an extended period.
}
\label{fig:s4}
\end{figure}

\clearpage
\newpage

\section*{Supplementary Figure S6. 
\\ Electric Potential of CytK M3IK from Molecular Dynamics}

\begin{figure}[hb]
    \centering
    \includegraphics[width=1.\linewidth]{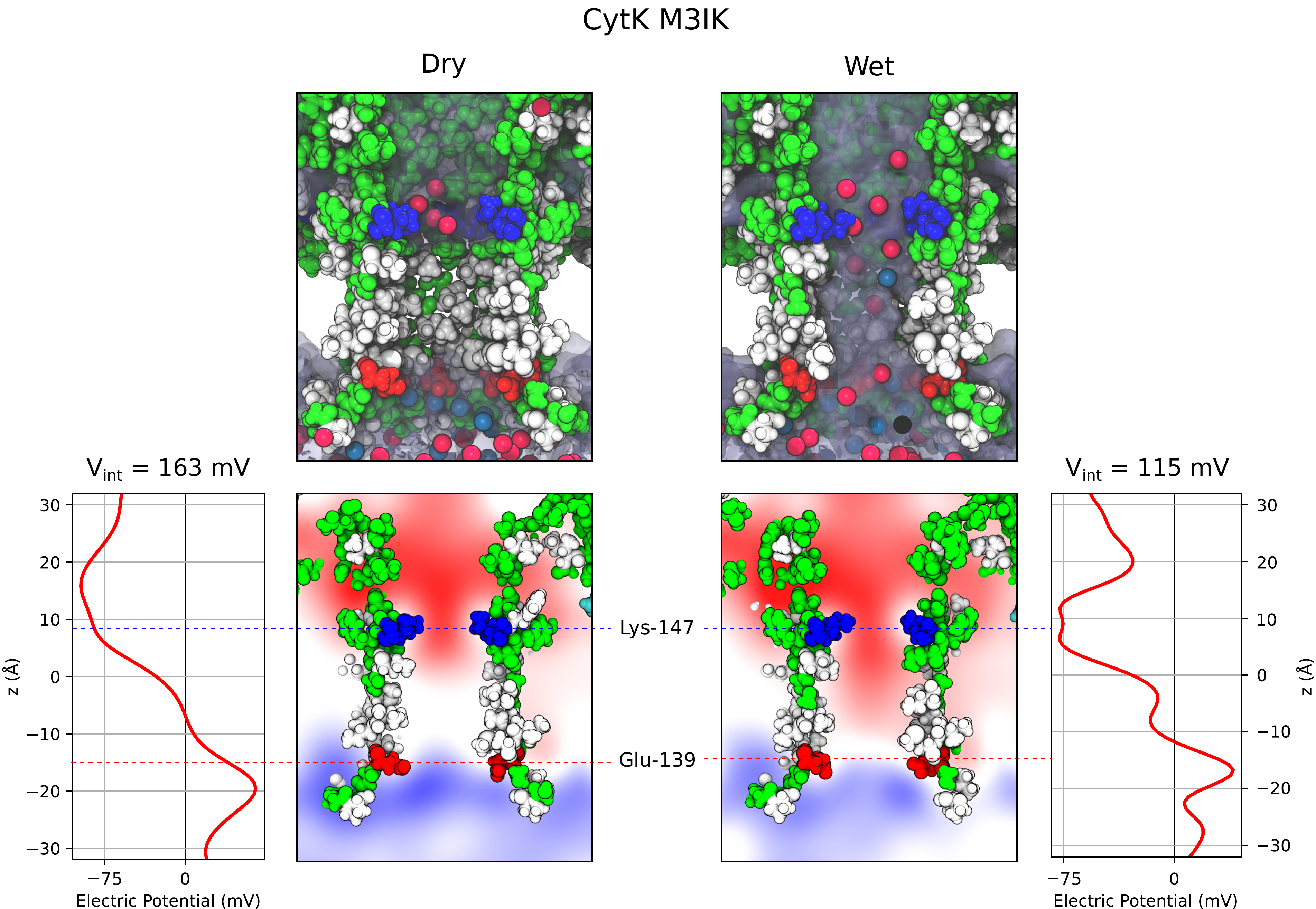}
\caption{
\textbf{Top. Lateral view of CytK M3IK in its dry and wet states.}
Water molecules are shown as transparent ice-blue surfaces, while protein and ions are displayed as Van der Waals spheres. 
Hydrophilic residues are colored green and hydrophobic residues white. Positively charged species (K$^+$ ions and Lys-147) are shown in blue, and negatively charged species (Cl$^-$ ions and Glu-139) in red.  The lipid membrane is omitted for clarity. 
\textbf{Bottom. Electrostatic potential obtained from equilibrium MD simulations}. Lateral graphs report the electric potential along the pore axis ($z$).
The potential, computed using the VMD \texttt{efpot} plugin, reveals an unexpected inversion of the intrinsic potential, with respect to the model pore in pure water.
Despite the arrangement of charged rings, with basic residues located toward the cis side and acidic ones toward the trans side, the resulting electrostatic potential exhibits the opposite polarity.
Supposedly, this inversion arises from the overscreening of counterions under conditions of high ionic strength (1\,M KCl), extreme confinement, and finite size effects, effectively reversing the intrinsic potential drop $V_\mathrm{int}$ within the channel. The inverted polarity results is in accordance with experimental measurements.
\label{fig:s6}
}
\end{figure}

\clearpage
\newpage
\section*{Supplementary Table S1. \\
CytK mutants sequences and primers.}
\textbf{CytK-WT}

\noindent
MAQTTSQVVTDIGQNAKTHTSYNTFNNEQADNMTMSLKVTFIDDPSADKQIAVINTTGSFMKANPTLSDAPVDGYPIPGASVTLRYPSQYDIAMNLQDNTSRFFHVAPTNAVE
ETTVTSSVSYQLGGSIKASVTPSGPSGESGATGQVTWSDSVSYK
QTSYKTNLIDQTNKHVKWNVFFNGYNNQNWGIYTRDSYHALYGNQLFMYSRTYPHETDARGNLVPMNDLPTLTNSGFSPGMIAVVISEKDTEQSSIQVAYTKHADDYTLRPGFTFGTGNWVGNNIKDVDQKTFNKSFVLDWKNKKLVEKKGSAHHHHHH

\vspace{1em}

\textbf{CytK-4D-S126I (CytK-S126I-K128D-Q145D-S151D-K155D)}

\noindent
MAQTTSQVVTDIGQNAKTHTSYNTFNNEQADNMTMSLKVTFIDDPSADKQIAVINTTGSFMKANPTLSDAPVDGYPIPGASVTLRYPSQYDIAMNLQDNTSRFFHVAPTNAVE
ETTVTSSVSYQLGGIIDASVTPSGPSGESGATGDVTWSDDVSYD
QTSYKTNLIDQTNKHVKWNVFFNGYNNQNWGIYTRDSYHALYGNQLFMYSRTYPHETDARGNLVPMNDLPTLTNSGFSPGMIAVVISEKDTEQSSIQVAYTKHADDYTLRPGFTFGTGNWVGNNIKDVDQKTFNKSFVLDWKNKKLVEKKGSAHHHHHH

\vspace{1em}

\textbf{CytK-S126I-K128M-T143I-Q145I-T147K}

\noindent
MAQTTSQVVTDIGQNAKTHTSYNTFNNEQADNMTMSLKVTFIDDPSADKQIAVINTTGSFMKANPTLSDAPVDGYPIPGASVTLRYPSQYDIAMNLQDNTSRFFHVAPTNAVE
ETTVTSSVSYQLGGIIMASVTPSGPSGESGAIGIVKWSDSVSYK
QTSYKTNLIDQTNKHVKWNVFFNGYNNQNWGIYTRDSYHALYGNQLFMYSRTYPHETDARGNLVPMNDLPTLTNSGFSPGMIAVVISEKDTEQSSIQVAYTKHADDYTLRPGFTFGTGNWVGNNIKDVDQKTFNKSFVLDWKNKKLVEKKGSAHHHHHH

Note: barrel domain are underlined and mutations with respect to the WT are in bold.\\

\vspace{0.3cm}
\textbf{Primers:}
\begin{table}[ht]
\centering
\begin{tabular}{ll}
\hline
\textbf{Primer name} & \textbf{Primer sequence} \\
\hline
T7p           & TAATACGACTCACTATAGGG \\
T7t           & GCTAGTTATTGCTCAGCGG \\
Insgen\_URv   & AGCAGCCAACUCAGCTTCCTTTCGGGCTTTG \\
Insgen\_UFw   & AAATAATTTUGTTTACTTTAAGAAGGAGATATAGCC \\
pT7gen\_UFw   & AGTTGGCTGCUGCCACCGCTGAGCAATAAC \\
pT7gen\_URv   & AAAATTATTUCTAGAGGGAAACCGTTGTGGTC \\
CK\_K128M\_Rv2 & CCGGACTCACCAGAAGGGCCGGACGGGGTCACGCTAGCCATGATGATACCACCCAACTGG \\
CK\_M3IK\_Fw    & GGCCCTTCTGGTGAGTCCGGTGCGATCGGTATTGTTAAATGGTCAGATTCCGTTAGCTAC \\
CK\_M2IFK\_Fw    & GGCCCTTCTGGTGAGTCCGGTGCGATCGGTTTCGTTAAATGGTCAGATTCCGTTAGCTAC \\
\hline
\end{tabular}
\caption{List of primers and their sequences.}
\label{tab:primers}
\end{table}

\end{document}